\documentclass[a4paper,11pt]{article}
\pdfoutput=1 

\usepackage{jcappub} 
\usepackage{lineno} 
\usepackage[T1]{fontenc} 
\usepackage{comment}
\usepackage{subcaption}
\usepackage{bm}
\usepackage[numbers]{natbib}
\usepackage{url}
\usepackage{hyperref}

\title{\boldmath \texttt{CosmoUiT}: A Vision Transformer-UNet Hybrid for Fast and Accurate Emulation of 21-cm Maps from the Epoch of Reionization}

\author[a,1]{Prasad Rajesh Posture,}
\author[a,2]{Yashrajsinh Mahida,}
\author[a]{Suman Majumdar,}
\author[a]{Leon Noble}

\affiliation[a]{Department of Astronomy, Astrophysics \& Space Engineering,\\ Indian Institute of Technology Indore,\\ Indore 453552, India}

\emailAdd{prasadposture121@gmail.com}
\emailAdd{yhmahida@gmail.com}
\emailAdd{mid.suman@gmail.com}
\emailAdd{leonnoblek@gmail.com}

\abstract{The observation of the redshifted 21-cm signal from the intergalactic medium will probe the epoch of reionization (EoR) with unprecedented detail. Various simulations are being developed and used to predict and understand the nature and morphology of this signal. However, these simulations are computationally very expensive and time-consuming to produce in large numbers. To overcome this problem, an efficient field-level emulator of this signal is required. However, the EoR 21-cm signal is highly non-Gaussian, making it challenging for neural networks to accurately capture the multi-scale correlations of the field. Here we introduce \texttt{CosmoUiT}, a UNet integrated vision transformer-based architecture, to overcome these difficulties. \texttt{CosmoUiT} emulates the 3D cubes of 21-cm signal from the EoR, for a given input dark matter density field, halo density field, and reionization parameters. \texttt{CosmoUiT} uses the multi-head self-attention mechanism of the transformer to capture the long-range dependencies and convolutional layers in the UNet to capture the small-scale variations in the target 21-cm field. Furthermore, the training of the emulator is conditioned on the input reionization parameters such that it gives a fast and accurate prediction of the 21-cm field for different sets of input reionization parameters. We evaluate the predictions of our emulator by comparing various statistics (e.g., bubble size distribution, power spectra) and morphological features of the emulated and simulated maps. We further demonstrate that this vision transformer-based architecture can emulate the entire 3D 21-cm signal cube with high accuracy at both large and small scales.}

\begin{document}
\maketitle
\flushbottom

\section{Introduction} \label{sec:intro}

The Epoch of Reionization (EoR) marks one of the major phase transitions in the history of the Universe. It corresponds to the phase when the UV radiation from the first luminous sources ionized the neutral hydrogen (HI) in the intergalactic medium (IGM). It is crucial to study the reionization process to understand the evolution of structures in the early Universe, yet our understanding is limited due to a lack of direct observations. Our current knowledge of the reionization process is based on the indirect probes, such as the Lyman-$\alpha$ forest \citep{Fan_2006, McGreer_2011, McGreer_2015}, the abundance of Lyman-$\alpha$ emitters \citep{Ouchi_2010, Robertson_2015, Zheng_2017}, and the Thomson scattering optical depth of the cosmic microwave background (CMB) \citep{Komatsu_2011, Planck_2016, Planck_2018}. However, these observations do not paint the complete picture of EoR, and to achieve that, we need the direct observations from the first ionizing sources and the evolution of the ionizing IGM.

The IGM is predominantly made up of neutral hydrogen. Therefore, 21-cm radiation emitted by the hyperfine spin-flip transition of neutral hydrogen acts as an excellent tracer of the evolution of IGM during EoR \citep{Furlanetto_2006, Pritchard_2012}. Many ongoing and upcoming radio interferometers, such as uGMRT \cite{Paciga_2013}, HERA\cite{HERA_2022a}, LOFAR \cite{Mertens_2025}, MWA \cite{MWA_2019}, and the upcoming Square Kilometer Array Observatory (SKAO) \cite{Koopmans_2015}, are making efforts to statistically detect this redshifted 21-cm signal coming from the EoR. These interferometers have already provided the upper limits on the spherical average power spectrum of the signal \cite{HERA_2025, MWA_2025, Gehlot_2019, Merterns_2020, HERA_2022a, HERA_2022b, Ceccotti_2025, Acharya_2024, Mertens_2025}. 

In the near future, SKAO is expected to provide the tomographic maps of the redshifted 21-cm signal \cite{Mellema_2015}, which will help us understand the morphology and evolution of ionized regions. To understand the effects of different astrophysical parameters and processes on the redshifted 21-cm signal, we simulate tomographic maps using radiative transfer or semi-numerical simulations. One of our primary goals is to estimate the reionization parameters from the observation of the 21-cm signal. Usually, to perform the Bayesian inference to estimate the reionization parameters, one needs to forward model this signal using the reionization simulations and then compute the signal statistics. However, these simulations are computationally very expensive to rerun for a large number of parameter sets, which is essential for Bayesian inference. One way to overcome this challenge is to use neural networks to emulate the statistics of the signal directly from the reionization parameters \cite{Schimt_2018, Jennigs_2019, Tiwari_2022, Sikder_2024, Breitman_2024, Mahida_2025}, bypassing the need to perform simulations. However, the 21-cm signal from the EoR is highly non-Gaussian; therefore, compressing it into a summary statistic will result in a loss of information. Since SKAO is expected to produce tomographic maps of the signal \cite{Mellema_2015}, we can directly use the signal maps to estimate the reionization parameters instead of compressing them into summary statistics. There are many works to estimate the reionization parameters directly from the tomographic maps using different neural network architectures \cite{Gillet_2019, Hortua_2020, Hassan_2020, Zhao_2022, Prelogovic_2022, Neutsch_2022, Schosser_2025, Ore_2025}; however, these neural networks also compress the signal into latent summary space, and this comes with the additional challenge of physical interpretation of these compressed summaries. Therefore, a more traditional approach would be field-level inference using Bayesian inference. To achieve this, we need an emulator that can emulate the entire 21-cm signal maps from the input reionization parameters, i.e, a field-level emulator.

Emulating a 3D 21-cm field is quite challenging. Recovering the morphological features at large and small scales in the emulated field is very difficult for any neural network architecture. Furthermore, the nonlinear evolution of the 21-cm signal, arising from the coupling of physical processes across different spatial scales, generates strong multiscale correlations and results in a highly non-Gaussian field. These correlations strongly influence how morphological features develop as reionization progresses. Therefore, one needs to develop a field emulator with an architecture capable of capturing these correlations within the field, which is a very complex task given the nature of the field. 

There are a few works in the literature that focus on emulating the HI signal simulation using neural networks \cite{Chardin_2019, Korber_2023, Zhao_2023, Hothi2025Reioinization, Mishra2025Reionization, Diao2025Reionization}. Two prominent efforts in this regard are \texttt{PINION} (Physics-Informed Neural Network for reIONization) \cite{Korber_2023} and \texttt{CRADLE} (Cosmological Reionization And Deep LEarning) \cite{Chardin_2019}. These two approaches of emulating the reionization field consider the IGM gas density and source field as inputs to predict the hydrogen ionization fraction or the time of the first reionization of each emulation pixel as the output. Both emulators suffer from distinct but important limitations. \texttt{PINION} reconstructs the ionization field by dividing the input gas and source density fields into sub-cubes and making localized predictions (cf. section 4.2 of \cite{Korber_2023}), which allows the model to capture small-scale features; however, since the entire field is not processed simultaneously, it struggles to capture large-scale correlations. \texttt{CRADLE}, on the other hand, is based on an autoencoder-style convolutional neural network that operates on 2D slices of the input fields, making independent predictions for each slice and combining them to construct the 3D field. Due to this slicing approach, it is difficult to capture the influence of ionizing sources across different slices. Although Gaussian kernel smoothing is applied to compensate for this effect, it leads to an underestimation of small-scale features in the output signal.

Therefore, neither of these models can simultaneously capture the large-scale and small-scale features of the reionization map and the inherent complex and time-evolving correlations between them. These emulators are trained on computationally expensive radiative transfer simulations; therefore, it is very difficult to generate a large training dataset for multiple reionization histories by changing the reionization parameters. As a result, these models are typically trained for a fixed reionization history. Moreover, since these architectures are designed to condition primarily on input fields rather than explicitly on reionization parameters, they are not suitable for field-level inference of reionization parameters without proper architectural modifications.

Motivated by the limitations of existing field-level emulators for the EoR 21-cm signal, we introduce \texttt{CosmoUiT}, a hybrid Vision Transformer (ViT) and UNet-based architecture to emulate the 3D cube of redshifted 21-cm signal from the EoR. The \texttt{CosmoUiT} takes the dark matter (DM) and halo fields along with the reionization parameters as its inputs and emulates the 3D ionization map of IGM, which is then converted to the brightness temperature map of the redshifted 21-cm signal. We aim to capture large as well as small-scale features in the emulated 3D maps; to achieve this, we have used the multi-head self-attention mechanism of the transformer to capture long-range dependencies, and convolutional layers in the UNet to capture small-scale variations in the field. Furthermore, we provide the reionization parameters during training to both the vision transformer block and UNet block to ensure that the reionization parameter influences the training to provide the output signal conditioned on the input reionization parameters. This hybrid design overcomes the limitations of earlier emulators. It provides a fast and accurate framework for generating 3D 21-cm redshifted signal cubes across the EoR parameter space, laying the groundwork for field-level inference with SKAO observations.

This article is organized as follows: In section~\ref{sec:simulations}, we describe the simulation framework and the reionization parameter space for the training and testing datasets. The section~\ref{sec:Methodology} presents the theoretical background of the vision transformer and a detailed description of \texttt{CosmoUiT} architecture. We present our results and validation against reference simulations in section~\ref{sec:Results}. Finally, section~\ref{sec:Summary} summarizes our findings and outlines future directions.

\section{Generation of Training Dataset} \label{sec:simulations}

We want our emulator to be able to predict the neutral hydrogen [$x_{\text{HI}}({\bf x})$] field given different sets of reionization (astrophysical) parameters. To achieve this, we generate a training dataset consisting of $\sim 7000$ simulations to train this emulator to learn the mapping from the 3D dark matter and halo fields, along with three input astrophysical parameters, to the corresponding 3D $x_{\text{HI}}$ field. For the development of this emulator, we focus on a single redshift and simulate coeval cubes centered at that redshift while varying the EoR parameters for the training dataset, as generating the 21-cm maps for a large number of astrophysical parameters at multiple redshifts is computationally very expensive. The process of simulating the 21-cm brightness temperature maps for the training dataset is described in the following paragraph.


The training dataset for our emulator was generated through a multi-step simulation process. First, we use a particle-mesh (PM) dark matter-only N-body simulation \cite{Bharadwaj_Srikant_2004, Mondal_2015} to generate the dark matter (DM) field at redshift $z=7$. The DM field was simulated on a $3072^3$ grid with a grid resolution of $0.07\, \rm{cMpc}$, giving us a simulation volume of $(215.04 \,  \rm{cMpc})^3$. We populate our simulation box with $1536^3$ dark matter particles, giving us particle-mass resolution of $1.09 \times 10^8 \rm{M_\odot}$. Then we use the Friends-of-Friends algorithm \cite{Davies_1985, Mondal_2015} to identify the halos in our DM density field. We use a fixed linking length of $0.2$ times the mean interparticle distance. The criterion for identifying the halo is that it should have at least $10$ DM particles to be considered as a halo, leading to a minimum halo mass of $1.09 \times 10^9 \rm{M_\odot}$ in our simulations.

Finally, we use seminumerical code \texttt{ReionYuga}\footnote{\url{https://github.com/rajeshmondal18/ReionYuga}} \cite{Choudhury_2009, Majumdar_2014, Mondal_2017} to generate the ionization field using the excursion set formalism \cite{Furlanetto_2004}. It takes the DM and Halo fields and generates the ionization fields. The code assumes that the hydrogen density follows the underlying DM density distribution and that the ionizing sources are hosted within the DM halos. \texttt{ReionYuga} has three free parameters: 1) \textbf{$\text{M}_{(h, min)}$}: This parameter sets the lower cutoff for the mass of the halo that participates in the reionization process. Only halos with the mass higher than $\text{M}_{(h, min)}$ have the sources that produce ionizing photons, 2) \textbf{$\text{N}_{ion}$}: We consider that the number of ionizing photons produced by the sources are proportional to the mass of the host DM halo and this dimensionless proportionality constant is $\text{N}_{ion}$. The $\text{N}_{ion}$ is a parameter that essentially quantifies the efficiency of the ionizing sources. 3) \textbf{$\text{R}_{mfp}$}: This parameter represents the mean free path of ionizing photons. For a more detailed discussion on this simulation, please refer to \cite{Choudhury_2009, Majumdar_2014, Mondal_2017}. We generate our training data set by varying these parameters in the following range - $\text{M}_{(h,min)} ( \times 10^9 \rm{M_\odot}) \in [10, 800]$, $\text{N}_{ion} \in [10, 200]$, and $\text{R}_{mfp}$ (Mpc) $\in [1.12, 40.32]$ and sample 7204 parameter sets from the uniformly gridded parameter space. 
\section{Vision Transformers} 
\label{sec:Methodology}

Transformer-based architectures have recently gained traction in cosmology as powerful tools for analyzing high-dimensional, non-Gaussian fields, owing to their ability to model long-range interactions through multi-head self-attention. These models have been successfully applied to large-scale structure and weak-lensing mass maps, where attention mechanisms improve the inference of cosmological parameters compared to purely convolutional approaches \cite{Hwang_2023, Kakadia_2025, Gondhalekar_2024}. In the context of radio cosmology, transformer-based frameworks have been explored for learning compact, information-rich summaries of 21-cm lightcones relevant for SKA-era observations \cite{Ore_2025}, as well as for foreground removal in ground-based CMB experiments \cite{Yan_foreground_2025}. Motivated by these advances, Vision Transformers \cite{Dosovitskiy_2020} offer a promising framework for the present work, as their self-attention mechanism enables modelling long-range correlations while naturally incorporating parameter conditioning. The following sections, therefore, describe the mathematical formalism of this framework and its application to the emulation task.

\subsection{Attention Mechanism}
Transformers were originally developed for natural language processing (NLP) tasks, where they effectively capture long-range dependencies using self-attention \cite{Vaswani_2017}. Vision Transformers (ViTs) extend this paradigm to visual data by representing images as sequences of patch embeddings \cite{Dosovitskiy_2020}. Since Transformers operate on 1D token sequences, images are partitioned into non-overlapping patches, flattened, and linearly projected into a latent embedding space.

In this work, we extend the ViT framework to 3D image cubes. Given an input volume $X \in \mathbb{R}^{h \times w \times d}$, the cube is divided into non-overlapping 3D patches of size $(P,P,P)$, resulting in $N = hwd/P^3$ patches. Each patch is flattened into a vector of dimension $P^3$, forming a sequence $\mathbf{X}_p \in \mathbb{R}^{N \times P^3}$. A learnable linear projection maps these patch vectors into a $D$-dimensional embedding space, yielding $\mathbf{X}_D \in \mathbb{R}^{N \times D}$, which serves as the input to the Transformer.

Self-attention enables each token to attend to all others in the sequence, allowing the model to capture both local and long-range dependencies. The input embeddings are linearly projected into queries, keys, and values using learnable weight matrices $\mathbf{W}^Q$, $\mathbf{W}^K$, and $\mathbf{W}^V$:
\begin{equation}
\mathbf{Q} = \mathbf{X}_D \mathbf{W}^Q, \quad
\mathbf{K} = \mathbf{X}_D \mathbf{W}^K, \quad
\mathbf{V} = \mathbf{X}_D \mathbf{W}^V.
\end{equation}
The attention operation for a single head is defined as
\begin{equation}
\text{Attention}(\mathbf{Q}, \mathbf{K}, \mathbf{V})
= \text{softmax}\left(\frac{\mathbf{Q}\mathbf{K}^T}{\sqrt{D_H}}\right)\mathbf{V},
\label{eq:attention_score}
\end{equation}
where $D_H$ denotes the hidden dimension.

To increase representational capacity, multi-head self-attention (MSA) is employed. For $h$ attention heads, the input is projected into $h$ sets of queries, keys, and values using independent projection matrices, with each head operating on a subspace of dimension $D_H = D/h$. The outputs from all heads are concatenated and linearly projected back to dimension $D$:
\begin{equation}
\text{MultiHead}(\mathbf{Q}, \mathbf{K}, \mathbf{V})
= [\mathbf{O}_1, \mathbf{O}_2, \dots, \mathbf{O}_h]\mathbf{W}^O .
\end{equation}

The MSA output is combined with the input via a residual connection followed by normalization. This is followed by a position-wise MLP, also wrapped with a residual connection and normalization, forming a standard Transformer encoder block.

\subsection{Architectural Strategies}
\label{sec:different_strategies}

The key limitations of \texttt{PINION} and \texttt{CRADLE} were the inability of the architectures to capture multi-scale dependencies and the use of a fixed set of astrophysical parameters. To address these limitations, we explored a series of architectural strategies that served as intermediate steps toward the final model. 

We tried a couple of different architecture strategies before developing \texttt{CosmoUiT} hybrid architecture. Mainly, architecture developed using transformer encoder layers with residual connections and transposed convolutions (referred to as \texttt{CosmoViT}) and one with an autoencoder-style CNN architecture (referred to as \texttt{CosmoUNet}). These forms of architecture were originally developed for image translation tasks. They rely on variations in the input to learn meaningful one-to-one mappings. However, in our case, the input fields (DM and halo fields) remained fixed for all output fields, and the information about the variation mainly came from the three reionization parameters, which the model couldn't capture properly. A detailed discussion of each architectural strategy, along with its model summaries, corresponding results, and metric scores, is presented in Appendix \ref{sec:architectural_strategies}. We learned from the shortcomings of these architectures and developed \texttt{CosmoUiT}, a hybrid architecture combining a vision transformer and a UNet.
\section{\texttt{CosmoUiT}: UNet integrated Transformer Emulator}\label{sec:uit_description}

In \texttt{CosmoUiT}, we introduced a transformer encoder layer before the UNet architecture. The transformer encoder enables the model to incorporate information about both the global field context and the reionization parameters before the data is passed through the UNet. This design integrates parameter specific variations directly into the feature representation, thereby enhancing the accuracy of mapping to the neutral hydrogen fraction field.

\subsection{Architecture}
\begin{figure}[htbp]
    \centering
    \includegraphics[width=0.98\textwidth]{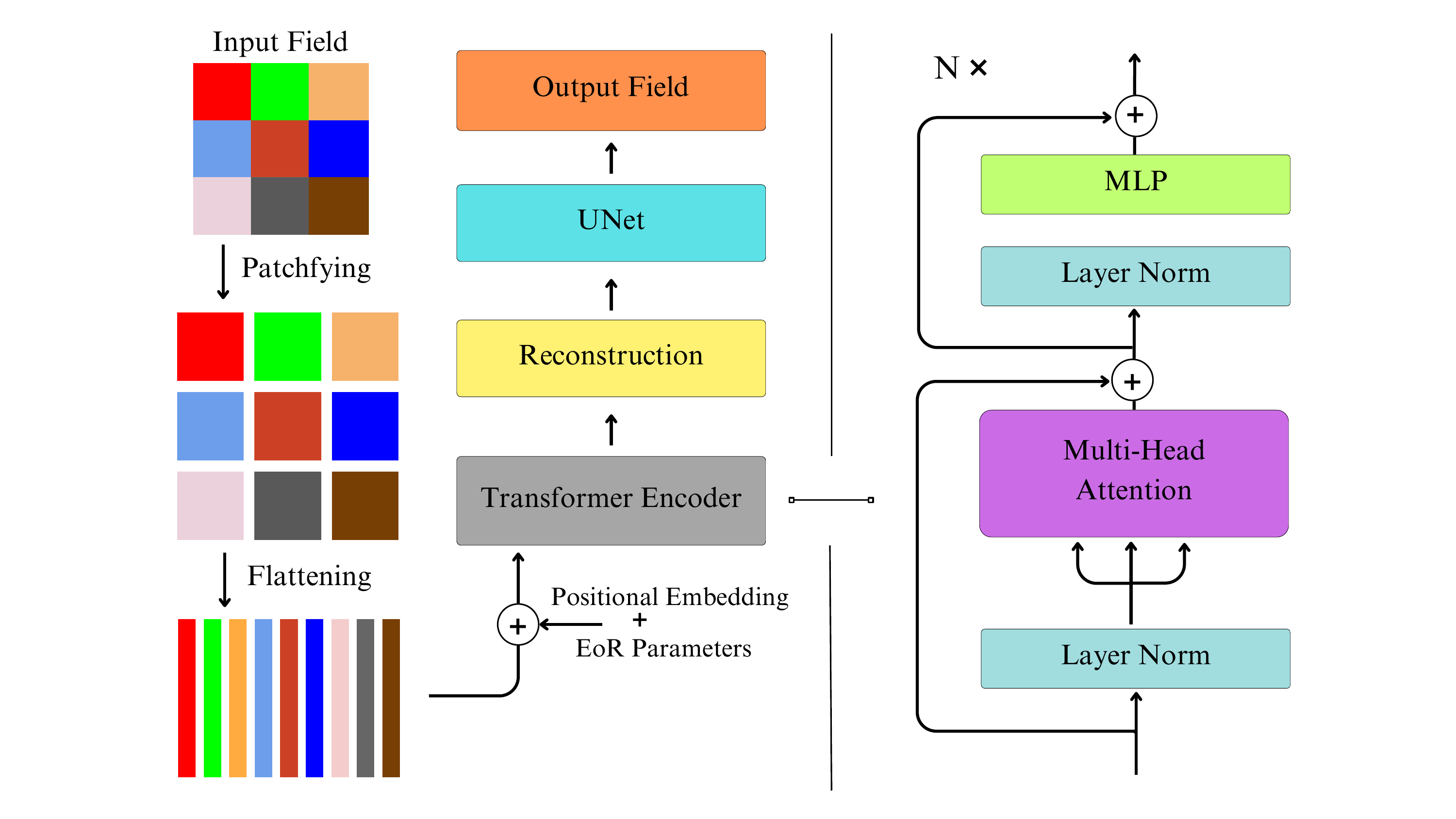} 
    \caption{Model Architecture of \texttt{CosmoUiT}.}
    \label{fig:CosmoUiTArchitecture}
\end{figure}

A detailed description of \texttt{CosmoUiT} architecture is illustrated in Figure \ref{fig:CosmoUiTArchitecture}. The process begins with the input fields, which are divided into small 3D subcubes called patches. Each patch contains \(8 \times 8 \times 8\) pixels, where each pixel corresponds to the size of  $2.24\, \rm{Mpc}$. Each patch is then flattened into a 1D array and projected into a lower-dimensional vector space. This step is known as tokenization. Transformers are permutation equivariant, meaning they do not inherently account for the order or position of tokens \cite{Xu_2023}. To address this, we explicitly add positional information \(P(i,j)\). This is based on two aspects: the position of each value within a token indicated by $i$; $i \in [0, d-1]$, where d is the embedding dimension, and the position of the token within the sequence denoted by $j$; $j \in [0, l-1]$, where l is the total number of tokens known as sequence length.

To condition the model on the three reionization parameters, we first project these parameters into the same dimensional space as the field tokens, and we call them parameter tokens. These parameter tokens are concatenated with the field tokens to form a unified input sequence. Inside the transformer encoder block, the sequence first passes through a layer normalization step. In the training of deep networks, the gradients of the backpropagation loss for each layer lead to increasingly smaller values as the depth of the network increases because the gradients of the backpropagated loss for each layer are calculated by multiplying the partial derivative of the loss of all the higher layers (chain rule of calculus). Thus, the network parameters are not updated effectively during the training. One solution to the vanishing gradient problem is to use skip connections in your network architecture. Therefore, in the transformer, we preserve an unnormalized copy of the input, which bypasses the multi-head self-attention block and is later used as a residual connection to help with gradient flow and mitigate the vanishing gradient problem. The normalized input is then used to compute three separate projections: Query ($\mathbf{Q}$), Key ($\mathbf{K}$), and Value ($\mathbf{V}$), which are used to estimate the attention scores following Equation \eqref{eq:attention_score}. After calculating attention scores, the resulting tokens are added to the preserved copy of the input, making each token aware of other tokens in the sequence. This makes the field tokens aware of the variation in the parameter tokens. The output of this attention block passes through another normalization layer and a multi-layer perceptron (MLP). The MLP typically consists of two linear layers with a non-linear activation function in between. Again, a residual connection is added using a preserved copy of the input before the MLP. This transformer encoder block is repeated $N$ times to extract complex interactions between the field and the parameters. After the final transformer encoder layer, only the field tokens are retained. These tokens are reshaped to reconstruct the spatial structure of the original field. This entire process is done separately for both the dark matter and halo fields.

\begin{figure}[htbp]
    \centering
    \includegraphics[width=0.98\textwidth]{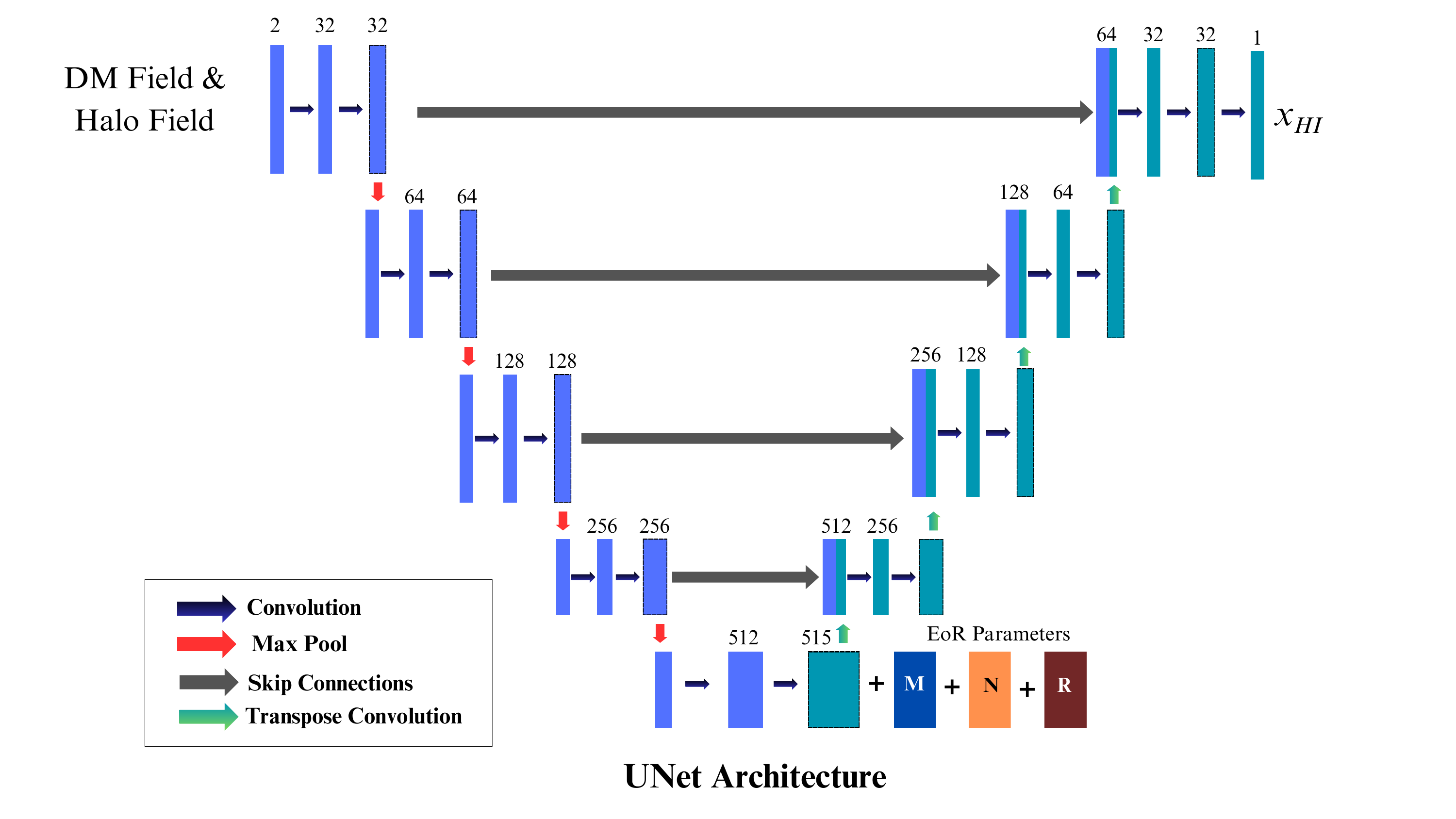} 
    \caption{Model Architecture of UNet.}
    \label{fig:CosmoUNetArchitecture}
\end{figure}

Once reconstructed, both the dark matter and halo fields are passed to a UNet architecture \cite{ROnneberger_2015}. This architecture consists of three main components: the encoder, the decoder, and the skip connections (cf. Figure \ref{fig:CosmoUNetArchitecture}). The encoder performs hierarchical feature extraction through successive convolutional and max-pooling layers. At each stage, the spatial resolution is reduced by half while the number of feature maps is doubled, starting from 32 maps at the input stage. This process continues until the bottleneck, where downsampling is no longer feasible. At this point, the three reionization parameters are incorporated by projecting them to the dimensionality of the bottleneck feature maps, enabling the network to condition on these parameters prior to upsampling. The decoder restores the output resolution by applying transpose convolutions. At each upsampling step, the output is concatenated with the corresponding encoder feature maps via skip connections, which transfer spatial information directly and help preserve small-scale features that would otherwise be lost during downsampling. The number of feature maps is halved while the spatial resolution is doubled at each stage. This is done till the neutral fraction field is reconstructed, having both spatial fidelity and parameter dependence. A summary of this architecture is provided in Table \ref{tab:cosmouit96_summary_corrected}.

\begin{table}[htbp]
\centering
\begin{tabular}{|l|c|c|c|}
\hline
\textbf{Component} & \textbf{No. of Feature Maps} & \textbf{Filter Size} & \textbf{Activation Function} \\ \hline
\textbf{Vision Transformer}      &                       &                           &                              \\ \hline
Patch Embedding       & $1728$ Tokens                    & $8\times8\times8 $                   & -                            \\ \hline
Projection Layer &   $512 \xrightarrow{} 512$      & -                         & -                            \\ \hline
Position Embedding & $(1, 1728, 512)$             &                          & -                            \\ \hline
Parameter Embedding    & $3 \xrightarrow{} 512  \xrightarrow{} 5120$                   & -                         & ReLU (inside)                            \\
 & $(n, 10, 512)$        &                          &                             \\ \hline
Transformer Encoder &      8Layers          &                          & ReLU (FFN)                         \\ \hline
\hspace{0.5cm} - Self-Attention  & 8 Heads              & -                         & -                            \\ 
 & $(n, 8, 1728+10, 64)$             &                       &                             \\ \hline

\hspace{0.5cm} - Feedforward     & $512$             & -                         & ReLU                         \\ \hline
\hspace{0.5cm} - Layer Normalization & 512                  & -                         & -                            \\ \hline
\textbf{UNet3D}      &      (Reconstructed Output)                 &                           &                              \\ \hline
Encoder Layer 1         & 32 Filters            & $3\times 3\times3$                     & LeakyReLU                         \\ \hline
Encoder Layer 2         & 64 Filters            & $3\times 3\times3$                     & LeakyReLU                         \\ \hline
Encoder Layer 3         & 128 Filters           & $3\times 3\times3$                     & LeakyReLU                         \\ \hline
Encoder Layer 4         & 256 Filters           & $3\times 3\times3$                     & LeakyReLU                         \\ \hline
Encoder Layer 5         & 512 Filters           & $3\times 3\times3$                     & LeakyReLU                         \\ \hline

Bottleneck + Parameters             &       512+3               &                           &                              \\ \hline
Skip+Decoder Layer 5      & 256 Filters           & $2\times2\times2$                    & LeakyReLU                            \\ \hline
Skip+Decoder Layer 4       & 128 Filters           & $2\times2\times2$                    & LeakyReLU                            \\ \hline
Skip+Decoder Layer 3     & 64 Filters            & $2\times2\times2$                     & LeakyReLU                            \\ \hline
Skip+Decoder Layer 2      & 32 Filters            & $2\times2\times2$                     & LeakyReLU                            \\ \hline
Decoder Layer 1     & 1 Filter            & $2\times2\times2$                   & Clamp[0,1]                            \\ \hline
\end{tabular}
\caption{Summary of the \texttt{CosmoUiT} architecture}
\label{tab:cosmouit96_summary_corrected}
\end{table}

\subsection{Training}
\begin{figure}[htbp]
    \centering
    \includegraphics[width=0.98\textwidth]{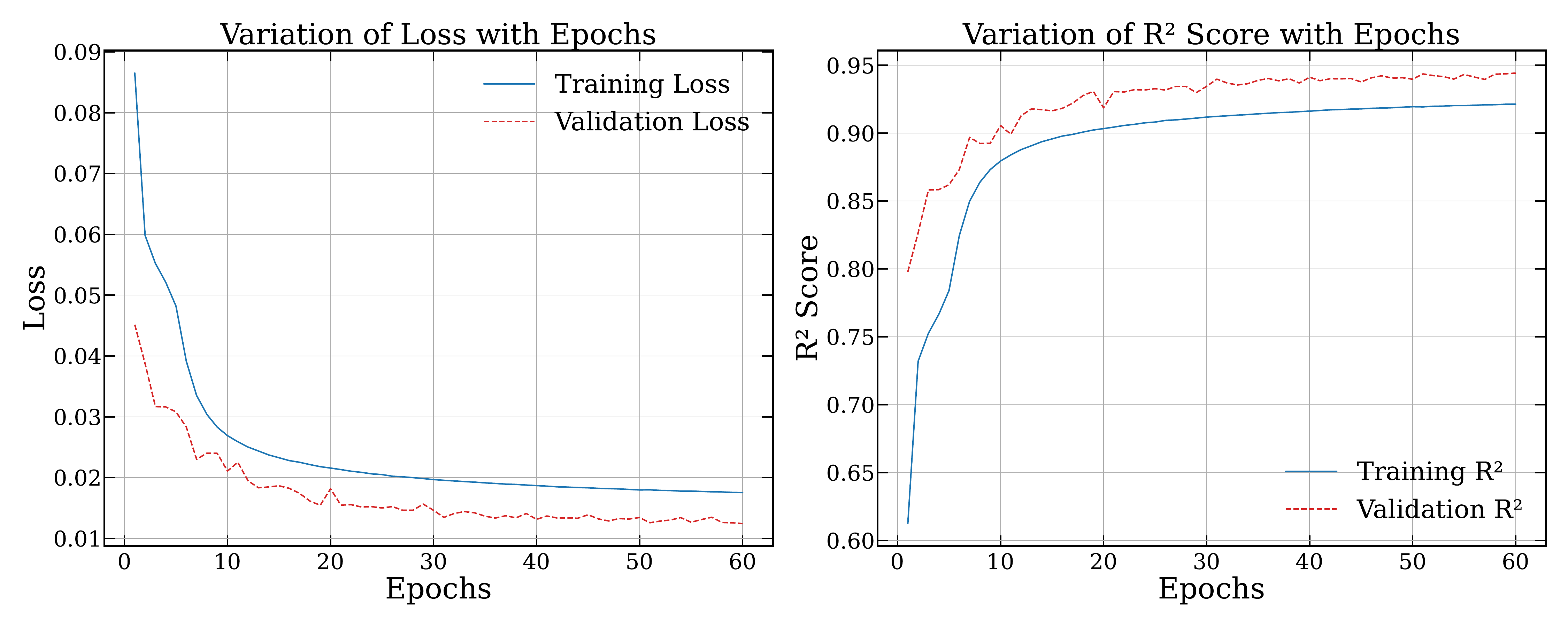} 
    \caption{Variation of loss and  ($\text{R}^2$) score for different epochs for training and validation data.}
    \label{fig:variation}
\end{figure}

We downsampled the training datasets from $384^3$ to $96^3$, so the memory requirements for training are within our available resources. To enhance generalization and avoid bias toward any specific spatial orientation, we applied data augmentation by incorporating all possible 3D orientations (rotations and reflections) of the input and corresponding output cubes. The dataset consists of $7204$ (parameter combinations) $\times~48$ (all possible orientations) of input-output pairs. It was split into $80\%$ training and $20\%$ validation subsets, ensuring coverage across the full range of reionization parameter values. The model was evaluated using mean squared error (MSE) and coefficient of determination ($\text{R}^2$) as performance metrics during training. We trained it for $60$ epochs using the Adam optimizer with a learning rate of $10^{-4}$. The training was conducted with a batch size of $16$ on an NVIDIA A100-SXM4-40GB GPU, consuming approximately $110$ GPU hours. The Figure \ref{fig:variation} shows the variation of MSE loss and $\text{R}^2$ score over the number of training epochs. The MSE loss for training and validation data decreases exponentially, implying that the model generalizes well for the unseen data. The model achieves a validation MSE of $0.012$ and an $\text{R}^2$ score of $0.94$. Once trained, the model requires $\sim 0.03$ seconds to generate a single 3D cube for a given set of reionization parameters on an RTX A4000 (16 GB) GPU. The inherently parallel nature of the model allows multiple cubes to be generated concurrently, depending on available GPU memory.

\subsection{Performance Metrics}
\label{sec:perform_metrics}

We use the following three metrics to evaluate the prediction from the emulator.

\begin{enumerate}
    \item Mean Squared Error (MSE):

        The MSE quantifies the average of the squared differences between predicted and true values. It penalizes larger errors more heavily and provides a direct measure of voxel-wise discrepancy:

        \begin{equation}
        \text{MSE} = \frac{1}{n} \sum_{i=1}^{n} (y_i - \hat{y}_i)^2
        \label{eq:mse}
        \end{equation}

    \item $\text{R}^2$ Score:

        The $\text{R}^2$ score evaluates the proportion of variance in the ground truth that is captured by the predictions. A score of 1 indicates perfect prediction, while a score of 0 corresponds to the performance of a naive mean predictor:

        \begin{equation}
        \text{R}^2 = 1 - \frac{\sum_{i=1}^{n} (y_i - \hat{y}_i)^2}{\sum_{i=1}^{n} (y_i - \bar{y})^2},
        \label{eq:r2score}
        \end{equation}
        
        Here, $y_i$ and $\hat{y}_i$ are the true values and predicted values, respectively, and $\bar{y}$ is the mean of the true values.

        \item Structure Similarity Index Measure (SSIM):

        The MSE and $\text{R}^2$ are metrics that compare voxel-wise distributions and do not compare the similarity of the spatial structure and morphology of the fields. We use SSIM to measure structural similarity between the emulated and simulated images. We use a sliding Gaussian window or a block window to produce an SSIM map from the entire image, and the SSIM value is obtained by averaging this map.  SSIM ranges from $-1$ to $1$, where values near $1$ indicate strong structural similarity, values around 0 indicate no similarity, and values close to $-1$ suggest strong anti-correlation. The SSIM is computed using the following equation:

        \begin{equation}
        \text{SSIM}(x, y) = \frac{(2\mu_x \mu_y + C_1)(2\sigma_{xy} + C_2)}{(\mu_x^2 + \mu_y^2 + C_1)(\sigma_x^2 + \sigma_y^2 + C_2)}~~,
        \label{eq:ssim}
        \end{equation}

        where $x$ and $y$ are the patches of the true and predicted fields, $\mu_x$ and $\mu_y$ are their respective means, $\sigma_x^2$ and $\sigma_y^2$ are the corresponding variances, $\sigma_{xy}$ is the covariance between the patches, and $C_1$ and $C_2$ are constants used to stabilize the division.

\end{enumerate}
\section{Results}\label{sec:Results}

We test the trained emulator on several cases spanning different choices of reionization parameter sets. These cases are drawn from an independent post-training test set of $\sim 700$ realizations, corresponding to $10\%$ of the total dataset, with the remaining $80\%$ and $10\%$ used for training and validation, respectively. For these sets of parameters, we first generate 3D $x_{\mathrm{HI}}$ fields and the corresponding 21-cm brightness temperature fields ($\delta T_b$). The emulator’s outputs are then compared against the corresponding simulation results using the performance metrics introduced in Sec \ref{sec:perform_metrics}. We also assess how well the emulator reproduces higher-level summary statistics of the target fields, such as the bubble size distribution and the power spectrum. Our final goal is to use \texttt{CosmoUiT} to do Bayesian field-level inference; however, to assess the statistical robustness of the fields predicted by \texttt{CosmoUiT}, we infer the reionization parameters through Bayesian inference considering the 21-cm power spectrum estimated from the predicted fields as a summary statistic. The details of the inference, along with the posterior plots, are presented in the appendix \ref{Parameter-inference}.

\subsection{\texorpdfstring{Comparison between $x_{\mathrm{HI}}$ Fields}{Comparison between x{\mathrm{HI} Fields}}}

\begin{figure}[htbp]
    \centering
    \includegraphics[width=0.97\textwidth]{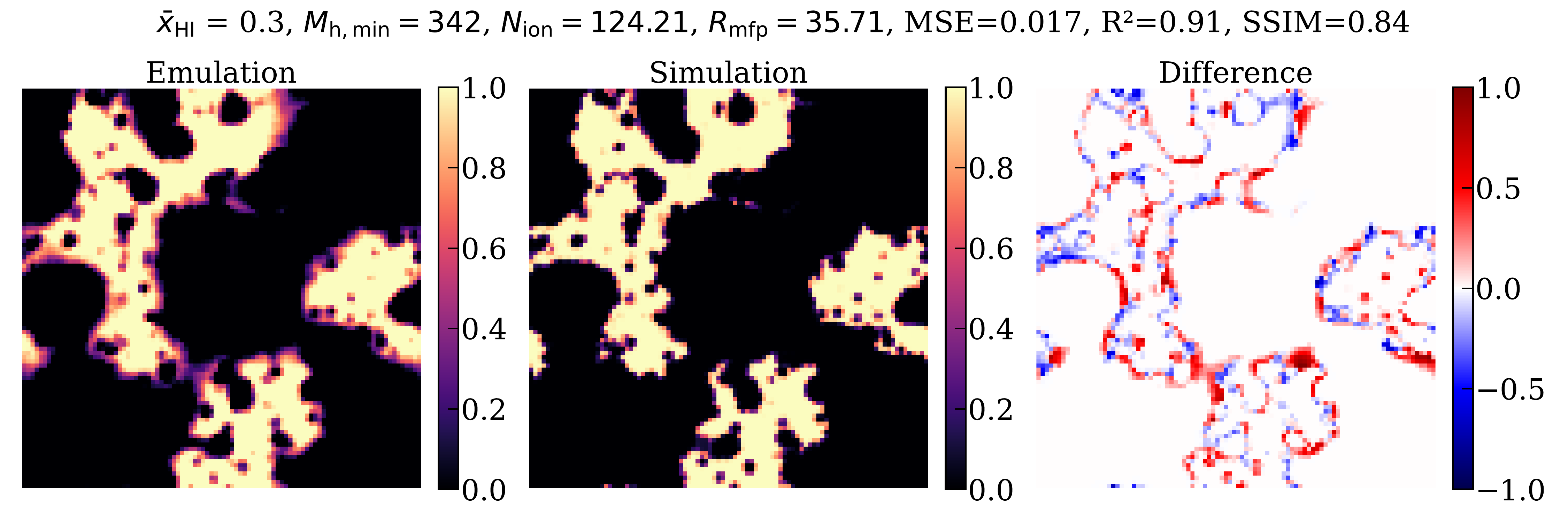}
    \vspace{0.5em}
    
    \includegraphics[width=0.97\textwidth]{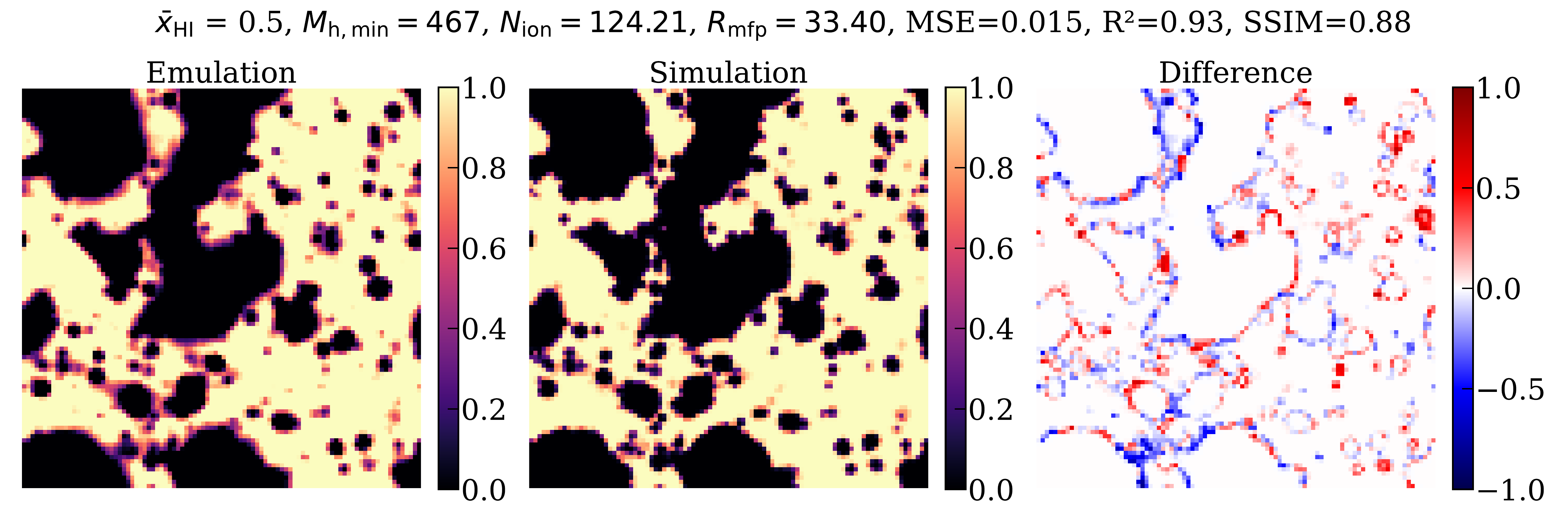}
    \vspace{0.5em}

    \includegraphics[width=0.97\textwidth]{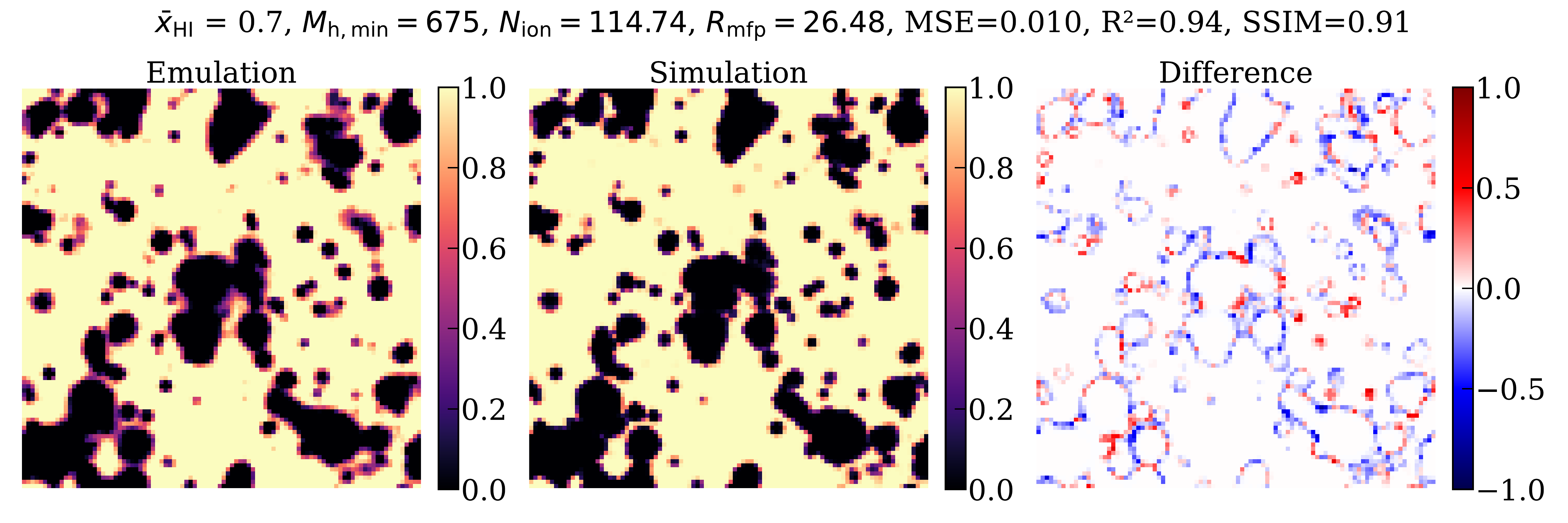}
    \caption{Comparison between \(x_{\mathrm{HI}}\) fields produced by emulation (\texttt{CosmoUiT}) and simulation (\texttt{ReionYuga}). In the first two columns, $1$ corresponds to neutral regions and $0$ corresponds to ionized regions. The third column gives the difference between these two fields. Each title contains the mass-averaged neutral fraction, the EoR parameter values in the units used for training, and the metric scores.}
    \label{fig:cosmouit96_predictions}
\end{figure}

\subsubsection{Performance Metrics Scores} \label{sec:performance_cosmouit96} 
Figure \ref{fig:cosmouit96_predictions} shows the middle slices of the 3D neutral fraction fields from \texttt{CosmoUiT} (emulated) and \texttt{ReionYuga} (simulated) outputs. Here, $0$ and $1$ denote ionized and neutral regions, respectively. The third column shows the difference between emulated and simulated fields. The three rows show results for three sets of reionization parameters. The corresponding parameter values, expressed in the units used as input to the network, are indicated in the title, along with the average mass-neutral hydrogen fraction ($\bar{x}_\mathrm{HI}$) and the associated performance metric scores. The visual comparison of the first two columns demonstrates good agreement on the large-scale structure and overall morphology of the ionized regions. Performance metrics also support this qualitative match. However, in the third column, where the difference between these two fields is plotted, it is evident that the primary source of errors in the predicted field is the boundary between the ionized and neutral regions. Since there is an abrupt, step-function-like change in the $x_\mathrm{HI}$ values at these boundaries, the emulator cannot capture it properly and instead predicts a gradual change \cite{Tsimenidis_2020}. As a result, the boundaries appear fuzzier than well-defined; hence, we refer to it as the fuzzy boundary problem. This problem also leads to an overestimation of bubble sizes when computing the bubble size distribution (Figure \ref{fig:bubble_size_distribution} and Table \ref{tab:peak_for_rmfp}).  Moreover, it contributes to the underprediction of small-scale fluctuations in the dimensionless power spectrum of the $x_{\mathrm{HI}}$ field (Figure \ref{fig:xHI_power_spectrum}). We try to quantify this error in prediction via uncertainty estimation (see Appendix \ref{sec:uncertainty_qantification}).

\subsubsection{Bubble Size Distribution} \label{sec:bubble_size_distribution}

In the EoR study, one of the quantities of great interest is the distribution of the size of the ionized regions or bubbles \cite{Furlanetto_2004, Giri_2018}. Several methods exist to quantify the bubble size distribution (BSD). In this work, we are using the mean free path method (MFP) \cite{Mesinger_2007}. This method gives the fraction of ionized bubbles in a given spherical-averaged size range. The BSD using the MFP method is estimated by randomly sampling ionized points in the ionization map and casting rays in random directions from each point until they hit a neutral region. Repeating this for many points builds a distribution of distances, which serves as a proxy for bubble sizes. This Monte Carlo-based MFP distribution is then convolved with a window function to correct for geometric biases. The corrected distribution gives a more accurate representation of the BSD during reionization \cite{Mesinger_2007, Lin_2016, Lu_2024}. We used the \texttt{Tools21cm}\footnote{\url{https://github.com/sambit-giri/tools21cm}} \cite{Giri_2018} Python package for computing the BSD.

\begin{figure}[htbp]
    \centering
    \includegraphics[width=0.65\textwidth, page=1]{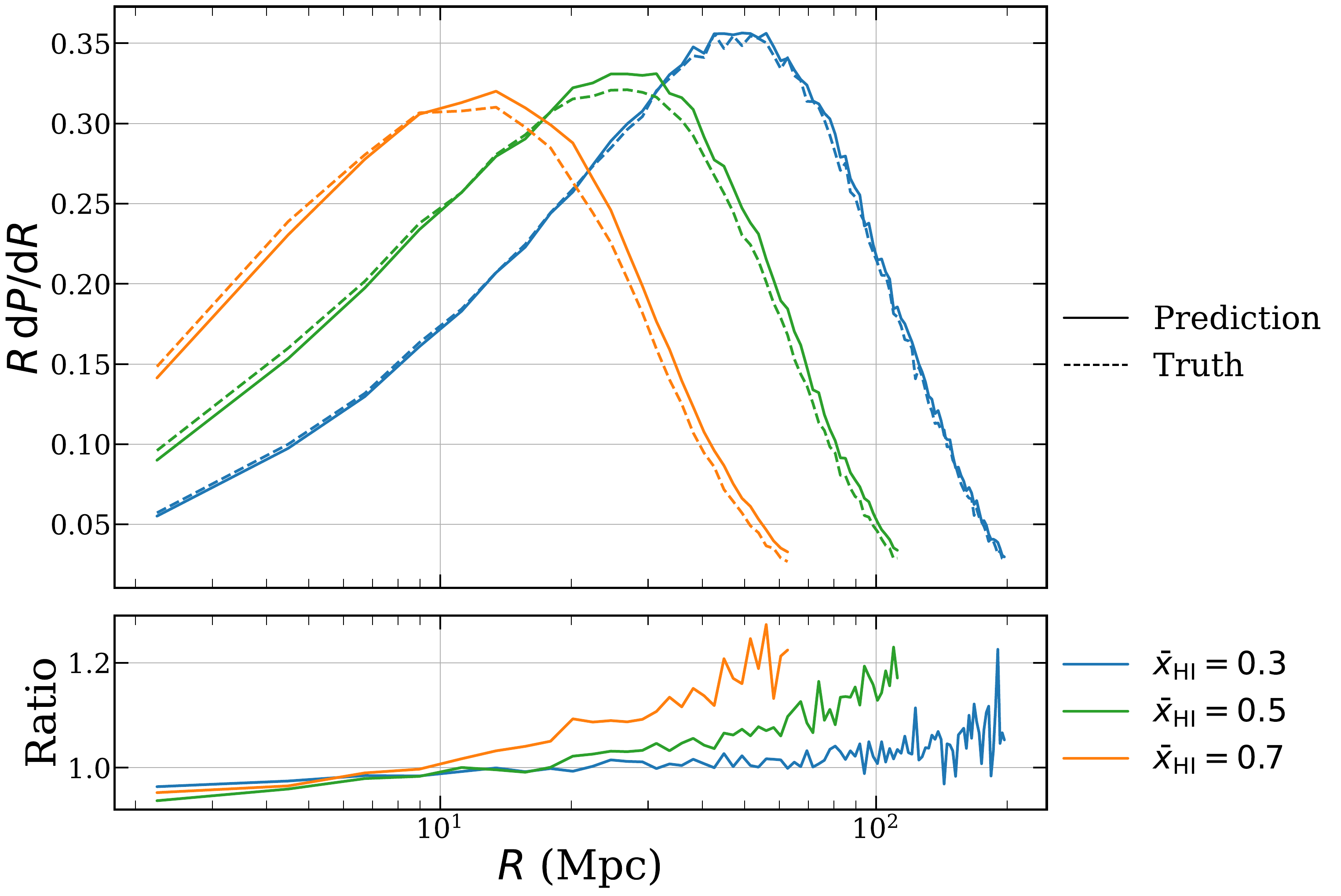} 
    \caption{\textit{Top Panel:} Bubble size distribution using the mean free path method. Solid lines are from emulation outputs, and the dashed lines are from simulation outputs. The blue, green, and orange colors correspond to the mass-averaged neutral fractions of $0.3$, $0.5$, and $0.7$, respectively. \textit{Bottom Panel:} The ratio plot is constructed by taking the ratio of the bubble size distribution obtained from the predicted field to that of the true field.}
    \label{fig:bubble_size_distribution}
\end{figure}

\begin{table}[h]
\centering
\begin{tabular}{|c|c|c|c|}
\hline
$\bm{\bar{x}}_{\mathrm{HI}}$ & \textbf{Emulation} & \textbf{Simulation} & $\bm{R}_{\mathrm{mfp}}$  \\
\hline
0.3 & 49.28 & 42.55 & 35.71 \\
0.5 & 31.36 & 26.88 & 33.40 \\
0.7 & 13.44 & 13.44 & 26.48 \\
\hline
\end{tabular}
\caption{Mean free path (in Mpc) at the peak of the bubble size distribution for different values of the volume average neutral hydrogen fraction $\bm{\bar{x}}_{\mathrm{HI}}$. The values are shown for both the emulated and simulated fields, along with the input mean free path parameter $\bm{R}_{\mathrm{mfp}}$ used in generating them.}
\label{tab:peak_for_rmfp}
\end{table}

Figure \ref{fig:bubble_size_distribution} presents the bubble size distribution (BSD) derived from both the emulation and the simulation output with solid and dashed lines, respectively. The top panel shows the absolute distributions, while the bottom panel illustrates the ratio of the predicted BSD to the simulated one. The blue, green, and orange colors correspond to volume average neutral hydrogen fractions of $0.3$, $0.5$, and $0.7$, respectively. Table \ref{tab:peak_for_rmfp} compares the mean free path corresponding to the BSD peak of the simulated and emulated fields, and the input mean free path parameter value used for generating these outputs for different mean neutral hydrogen fractions.

We observe that the bubble size distribution almost follows a normal distribution. The ratio plot in the bottom panel reveals that the BSD is increasingly overestimated after reaching and surpassing the peak, particularly in highly ionized fields. This overestimation arises due to the fuzzy boundary problem discussed in Section \ref{sec:performance_cosmouit96}. In \texttt{Tools21cm}, any cell with an ionization fraction greater than $0.5$ is considered ionized; hence, fuzzy boundaries are also considered to be ionized, leading to such an overestimation. Additionally, the offset between the BSD peak and the input mean free path $R_{\mathrm{mfp}}$ (see Table \ref{tab:peak_for_rmfp}) arises because $R_{\mathrm{mfp}}$ is a model parameter that acts as an upper limit on the distance ionizing photons can travel, rather than directly determining the typical bubble size. In the highly ionized fields, the $R_{\mathrm{mfp}}$ is smaller due to merging of bubbles. The peak of the BSD shows the typical size of ionized bubbles, which depends on both the mean free path $R_{\mathrm{mfp}}$ and the ionizing efficiency $N_{\mathrm{ion}}$. Because of degeneracies between these parameters, the bubble sizes can shift and may not directly reflect the input value of $R_{\mathrm{mfp}}$.

\subsubsection{Power Spectrum}



\begin{figure}[htbp]
    \centering
    \includegraphics[width=0.65\textwidth, page=1]{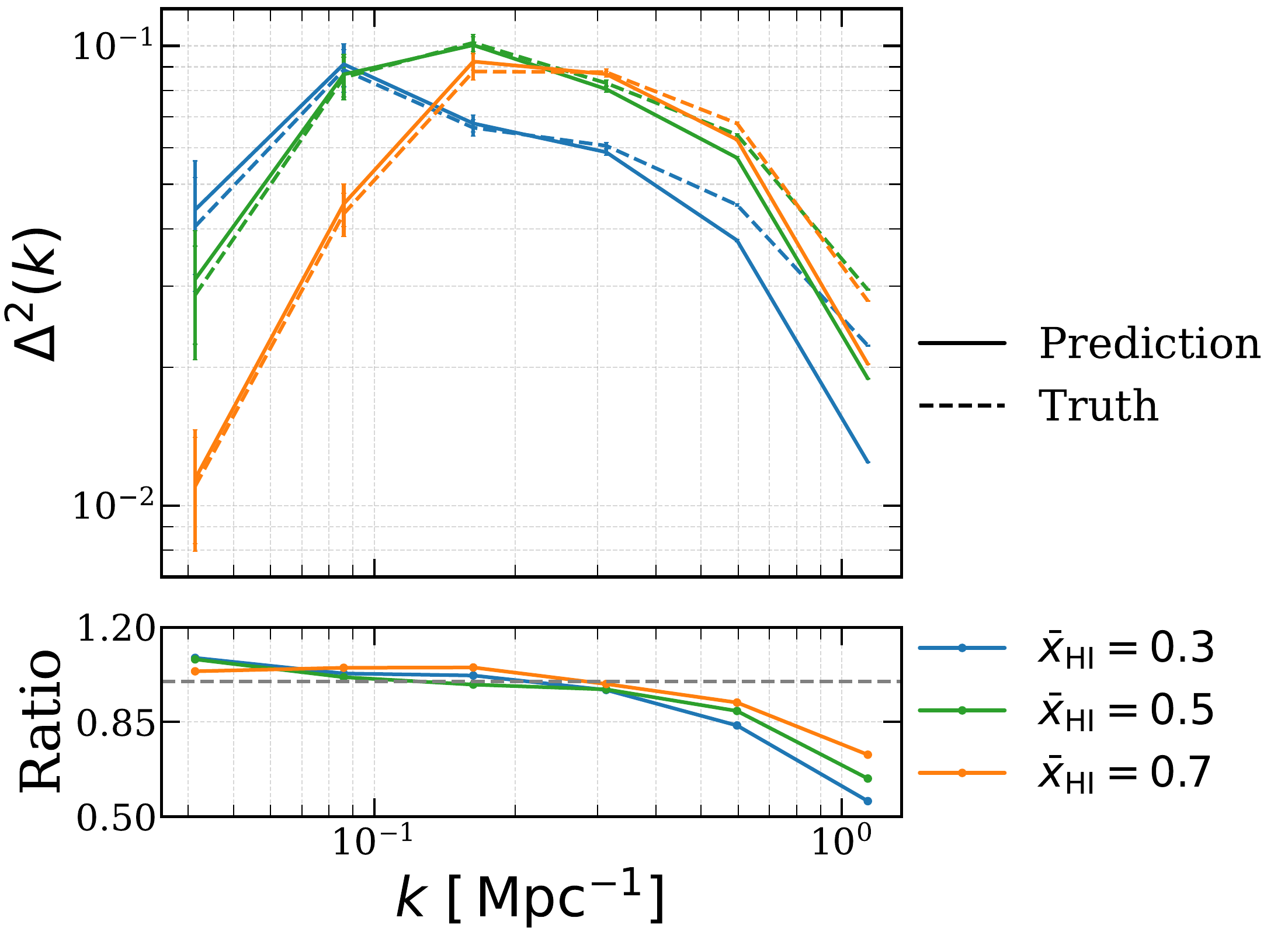} 
    \caption{\textit{Top Panel:} Dimensionless power spectrum of $x_{\mathrm{HI}}$ field with for varying volume average neutral fraction values. The solid lines correspond to \texttt{CosmoUiT} predictions, and the dashed lines correspond to \texttt{ReionYuga} simulations. The blue, green, and orange lines correspond to the mass-averaged neutral fraction of $0.3$, $0.5$, and $0.7$, respectively. \textit{Bottom Panel:}The ratio plot is constructed by taking the ratio of the dimensionless power spectrum from the predicted field to that of the true field.}
    \label{fig:xHI_power_spectrum}
\end{figure}

To evaluate the statistical properties of the emulated $x_{\mathrm{HI}}$ field, we compare its power spectrum against that of the simulated field. Following the approach in previous sections, we examine three values of the mass-averaged neutral fraction, as shown in Figure \ref{fig:xHI_power_spectrum}.

The power spectrum is obtained from the Fourier transform of the ionization field:
\begin{equation}
\delta(\vec{k}) = \int x_{HI}(\vec{r})\,e^{-2\pi i \vec{k} \cdot \vec{r}}\,d\vec{r},
\label{eq:xhi_fourier_transformer}
\end{equation}
where $\vec{r}$ and $\vec{k}$ denote spatial and Fourier coordinates, respectively. Taking the ensemble average:
\begin{equation}
\left\langle \tilde{\delta}(\vec{k}) \tilde{\delta}(\vec{k}') \right\rangle = (2\pi)^3 \delta^D(\vec{k} - \vec{k}')P(k),
\label{eq:xhi_ensemble_avg}
\end{equation}
where $P(k)$ is the spherically averaged power spectrum and $\delta^D$ is the Dirac delta function. We further define the dimensionless power spectrum as
\begin{equation}
\Delta^2(k) = \frac{k^3}{2\pi^2} P(k),
\label{eq:normalized_ps}
\end{equation}

Figure \ref{fig:xHI_power_spectrum} shows a comparison of the dimensionless power spectrum of the emulated and simulated $x_{\mathrm{HI}}$ fields. The top panel shows the power spectrum for three volume average neutral hydrogen fractions: $0.3$ (blue), $0.5$ (green), and $0.7$ (orange). The solid and dashed lines show the power spectrum of the emulated fields and simulated fields, respectively. The bottom panel displays the ratio of the predicted to the simulated power spectrum.  The outputs of emulation and simulation are in excellent agreement over a wide range of scales. The shape and amplitude of the predicted power spectrum closely follow that of the simulation, particularly at large length scales ($k \leq 0.3\,\mathrm{Mpc}^{-1}$), where the power of the predicted field falls within the error bars of the simulated field. It indicates that the \texttt{CosmoUiT} effectively captures the large-scale variations. At small length scales ($k > 0.3\,\mathrm{Mpc}^{-1}$), we observe a mild underprediction of power in the emulated fields, but less than 1 order of magnitude. This discrepancy becomes more evident at lower neutral fractions. The main reason is the fuzzy boundary problem. The emulator tends to smooth out the sharp ionization fronts that are present in the simulations. This smoothing creates more gradual variations at small scales. As a result, the emulator underpredicts the power on those scales. The emulator performs best for higher neutral fractions ($\bar{x}_{\mathrm{HI}} \approx 0.7$), where the ionized regions are relatively sparse and isolated, making them easier to reproduce. For highly ionized fields, the ionization morphology becomes complex due to percolation between ionized regions, and it is harder for the emulator to reproduce.

Compared to existing emulators, \texttt{CosmoUiT} gives a more balanced performance across scales. \texttt{CRADLE} captures large-scale features well but underpredicts small-scale power by up to an order of magnitude \cite{Chardin_2019}. \texttt{PINION}, on the other hand, recovers the small-scale fluctuations but misses the large-scale structure \cite{Korber_2023}. Our emulator bridges this gap by reproducing large-scale behaviour while maintaining reasonable agreement at smaller scales.

\subsection{\texorpdfstring{Comparison between $\delta T_b$ Fields}{Comparison between  
 delta Tb Fields}}
 



\begin{figure}[htbp]
    \centering
    \includegraphics[width=0.97\textwidth]{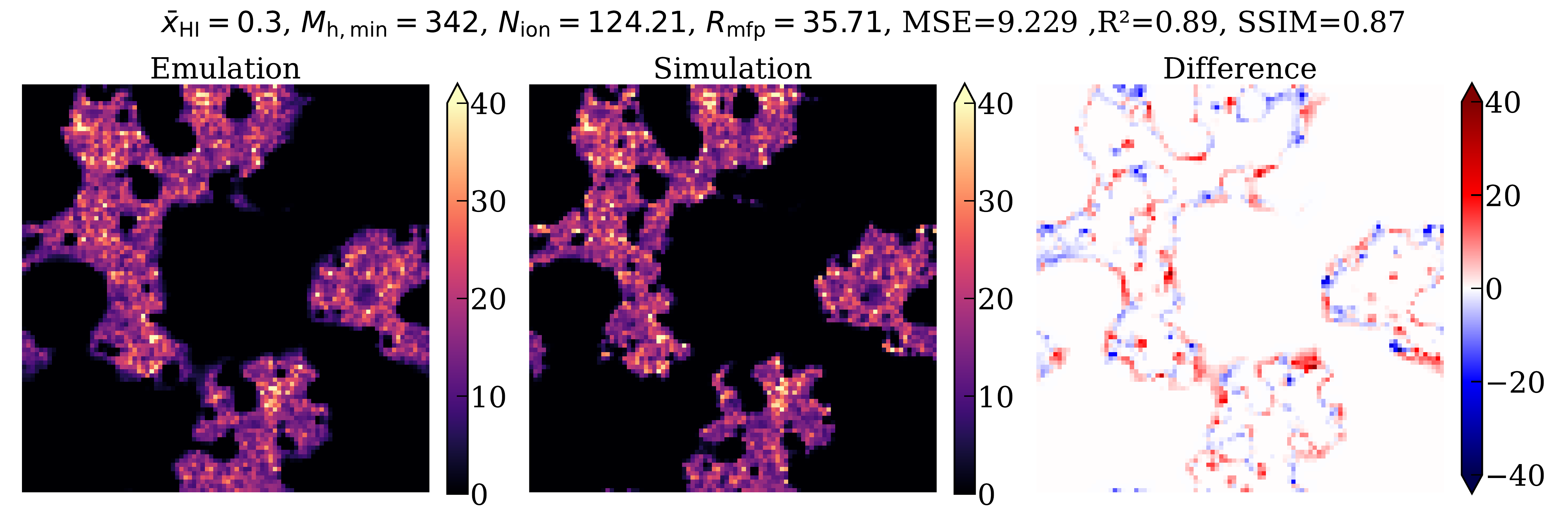}
    \vspace{0.5em}
    
    \includegraphics[width=0.97\textwidth]{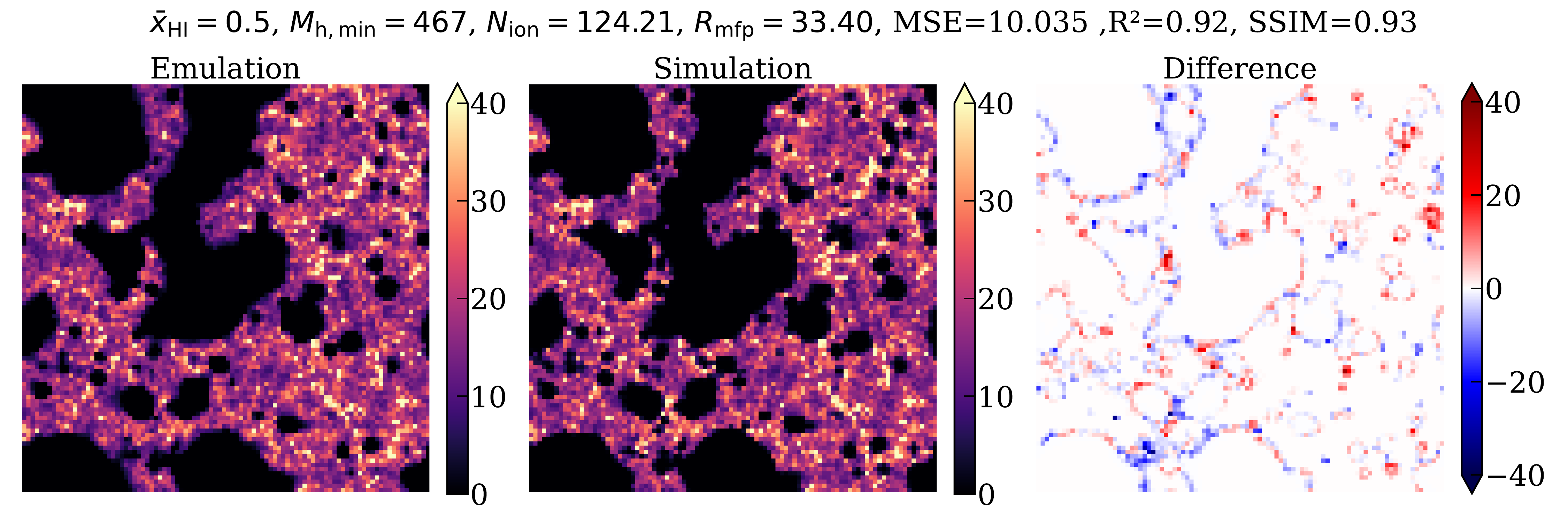}
    \vspace{0.5em}
    
    \includegraphics[width=0.97\textwidth]{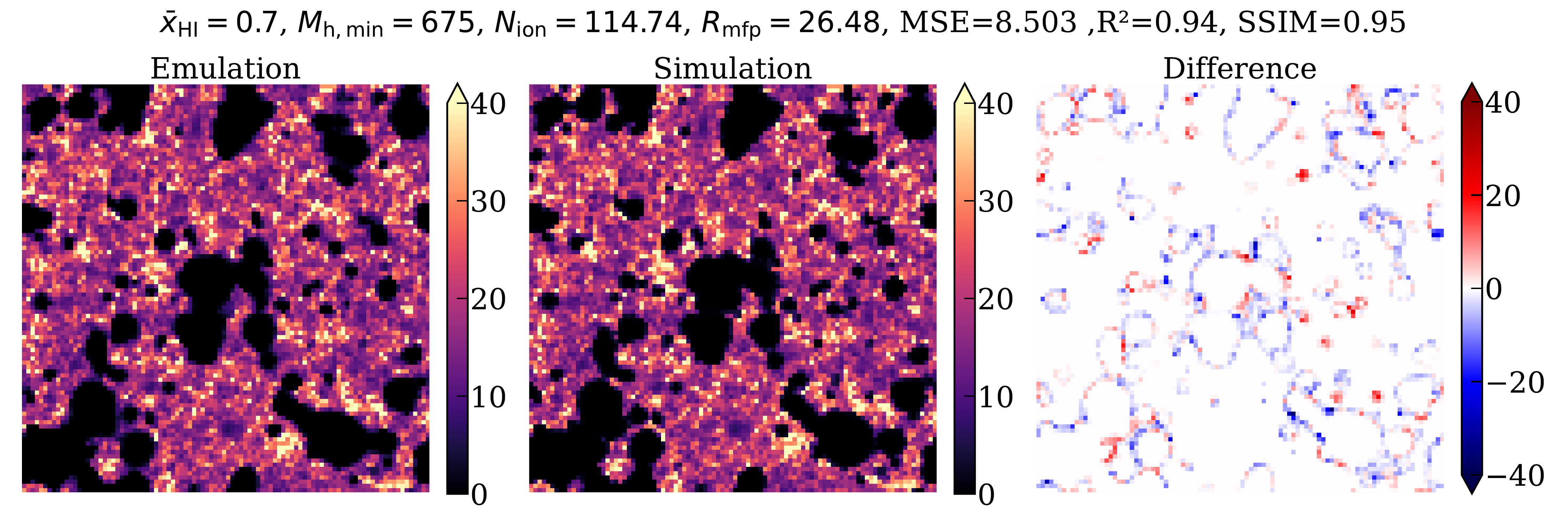}
    
    \caption{Comparison between 21-cm brightness temperature fields (in mK) produced by emulation and simulation. The first column contains the emulation output, the second the simulation output, and the third the difference between the two. Each title contains the mass-average hydrogen-neutral fraction, the EoR parameter values, and the metric scores.
    }
    \label{fig:prediction_21cm_all}
\end{figure}

\subsubsection{Performance Metrics Scores}


We calculate the same set of performance metrics for the emulated $\delta T_b$ fields as for the $x_{\mathrm{HI}}$ fields. Visual inspection of Figure \ref{fig:prediction_21cm_all} shows close agreement between emulated and simulated maps across all parameter sets, particularly in the morphology of ionized and neutral regions. Quantitatively, the  $\text{R}^2$ and SSIM values for the 21-cm fields exceed those obtained for the $x_{\mathrm{HI}}$ fields. This improvement arises due to the construction of the brightness temperature field: $\delta T_b$ is obtained by multiplying the neutral fraction field by the baryonic overdensity factor $(1 + \delta_b)$. Most prediction errors come from ionization boundaries, which often occur in low-density regions \cite{Bolton_2007, Mao_2009}. Since these low-density regions contribute less to the overall signal, the effect of boundary errors is reduced in the brightness temperature field, leading to higher metric scores compared to that of the $x_{\mathrm{HI}}$ fields.

\subsubsection{Power Spectrum}
The power spectrum of the redshifted 21-cm signal from EoR is the key observable for radio interferometers. We computed the spherically averaged 1D power spectrum for the 21-cm brightness temperature fields using the same Fourier formalism as described in Equations \eqref{eq:xhi_fourier_transformer}, \eqref{eq:xhi_ensemble_avg}, and \eqref{eq:normalized_ps}, replacing the $x_{\mathrm{HI}}$ field with the 21-cm brightness temperature field.

\begin{figure}[htbp]
    \centering
    \includegraphics[width=0.65\textwidth, page=1]{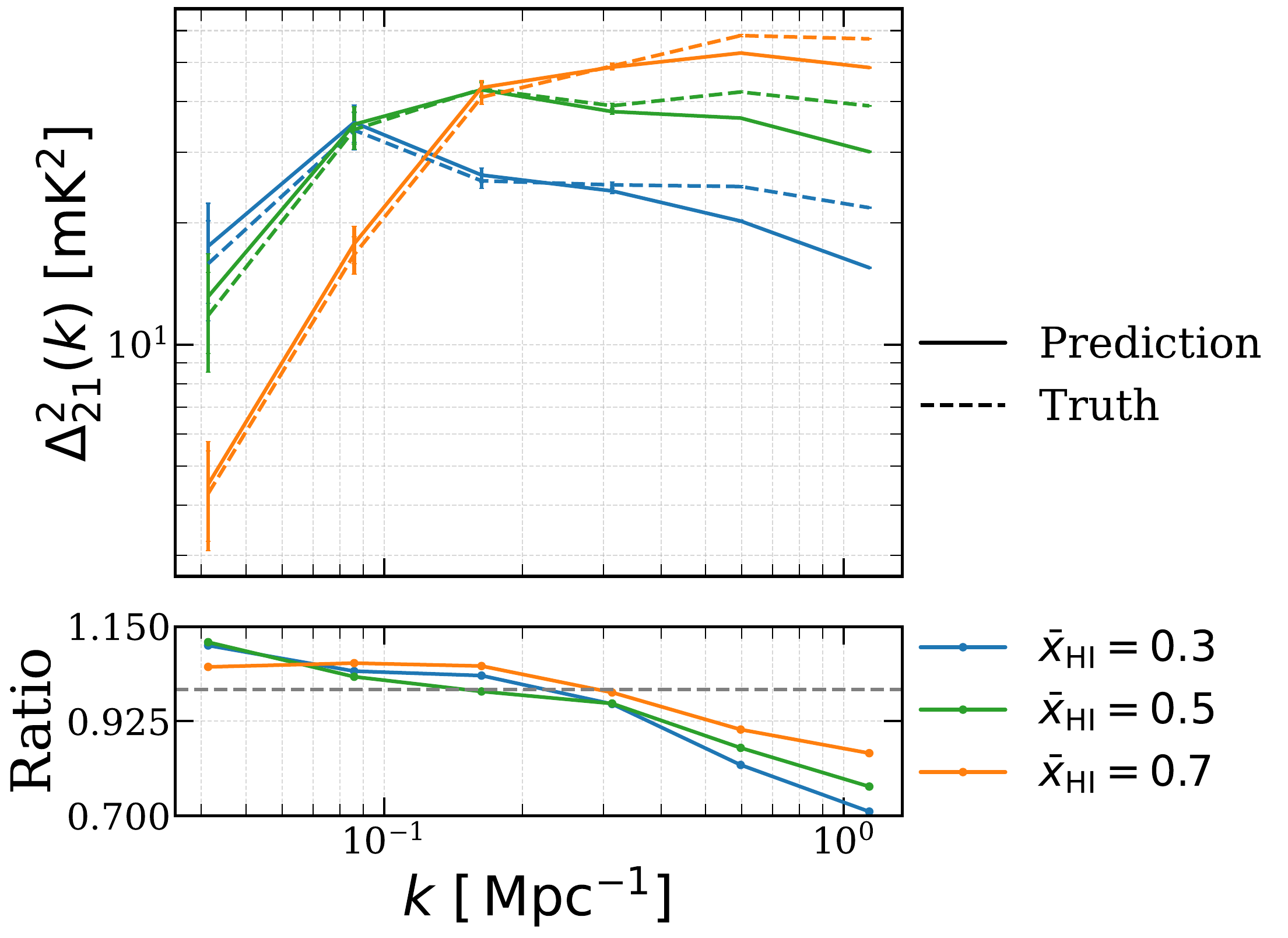} 
       \caption{\textit{Top Panel:} Dimensionless power spectrum of 21-cm brightness temperature field for varying mean hydrogen neutral fraction values. The solid lines correspond to \texttt{CosmoUiT} predictions, and the dashed lines correspond to simulation output. The blue, green, and orange lines correspond to the mass-average neutral fractions of $0.3$, $0.5$, and $0.7$, respectively. \textit{Bottom Panel:} Ratio plot for the above dimensionless power spectrum. The ratio is obtained by dividing the power spectrum of the predicted field by that of the true field.}
       \label{fig:21cmBT_power_spectrum}
\end{figure}
Figure \ref{fig:21cmBT_power_spectrum} demonstrates that the \texttt{CosmoUiT} predictions match the simulation outputs closely (within the sample variance limit) across all scales. Compared to the $x_{\mathrm{HI}}$ field power spectrum, the 21-cm power spectrum shows reduced errors for higher $k$-modes. This improvement stems from small-scale fluctuations in the density field. Unlike the ionization field, these gravitational fluctuations are directly passed as inputs to the model. Therefore, they are accurately represented in the predicted brightness temperature field. This means that, even if the model introduces some errors at ionization boundaries, the presence of realistic small-scale fluctuations from the density field helps preserve the statistical structure of the 21-cm signal. Consequently, when ensemble averaged in the computation of the power spectrum, the correctly captured small-scale modes compensate for localized boundary errors, resulting in improved agreement with the simulation across a wide range of scales, particularly at higher $k$-modes.

\subsection{Out-of-Domain (OOD) Generalization}
A key requirement for any robust emulator is the ability to generalize beyond the specific samples seen during its training. In the context of cosmological emulation, this refers to the model's capacity to make accurate predictions even when the input fields are generated with different initial random seeds than those used during training. This capability is known as out-of-domain generalization.

In our setup, the emulator was trained using a single realization of the dark matter and halo fields, with variability across training samples coming solely from the three reionization parameters. Since the spatial structure of the input fields was fixed during training, a model that overfits to these configurations rather than learning the underlying mapping would likely fail to generalize to new realizations. To test the generalization ability of \texttt{CosmoUiT}, we evaluated it on inputs generated using entirely different random seeds than those used in the training data. These new realizations contain different spatial configurations of the dark matter and halo fields, which the model has not seen during training.

\begin{figure}[htbp]
    \centering
    \includegraphics[width=0.97\textwidth]{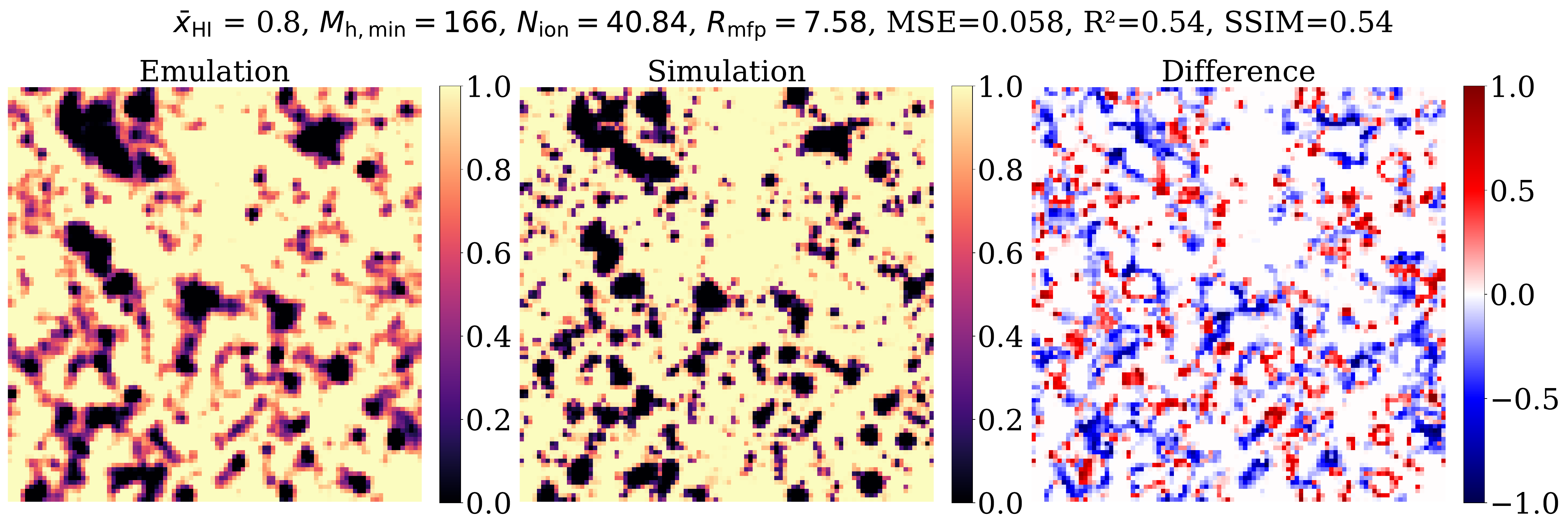}
    \vspace{0.5em}
    \includegraphics[width=0.97\textwidth]{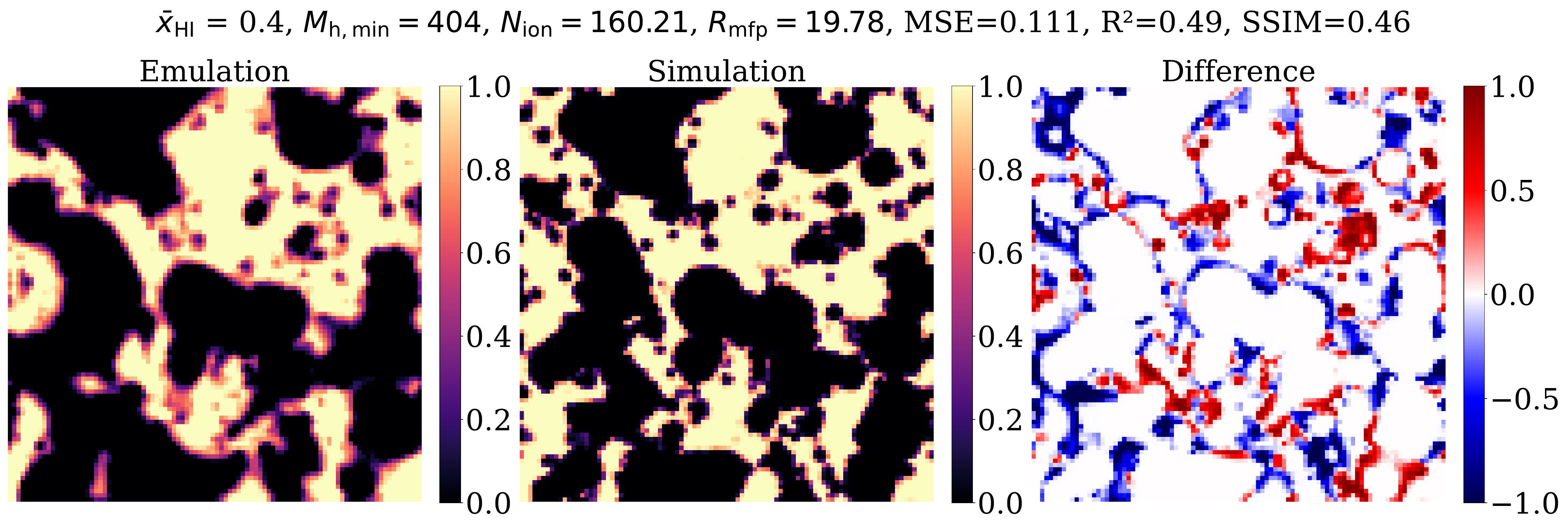}

    \caption{   Comparison between \(x_{\mathrm{HI}}\) fields produced by emulation (\texttt{CosmoUiT}) and simulation (\texttt{ReionYuga}) for input fields generated using random seeds not used in training. In the first two columns, $1$ corresponds to neutral regions and $0$ corresponds to ionized regions. The third column gives the difference between these two fields. Each title contains the mass-averaged neutral fraction, the EoR parameter values in the units used for feeding them to the architecture, and the metric scores.}
    \label{fig: prediction_unseen_rs}
\end{figure}


Figure \ref{fig: prediction_unseen_rs} shows examples of the predicted and simulated neutral fraction fields for these unseen realizations, along with the corresponding difference fields and metric scores. Despite the unfamiliarity of the input structures, our emulator can recover the overall ionization morphology and spatial features with impressive accuracy. The difference fields highlight the nature of the error. For seen input realizations, most of these errors were concentrated along the ionization boundaries. In contrast, for unseen input realizations, the errors are not confined to the boundaries, but also appear within the ionized and neutral regions, arising from over- and under-prediction of the neutral fraction. Due to this, we obtain higher MSE and lower  $\text{R}^2$ and SSIM compared to the seen realization cases. For the given examples, the $\text{R}^2$ score and the SSIM have approximately halved, while the MSE has increased by nearly a factor of six.

\begin{figure}[htbp]
    \centering
    \includegraphics[width=0.97\textwidth]{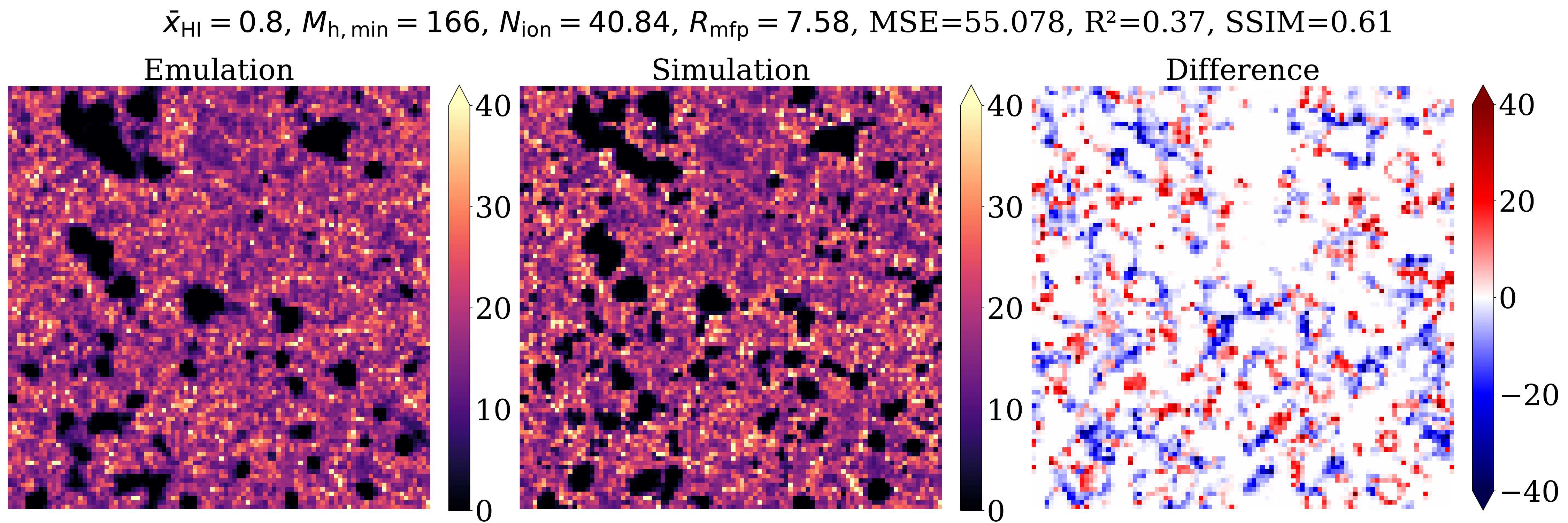}
    \vspace{0.5em}
    \includegraphics[width=0.97\textwidth]{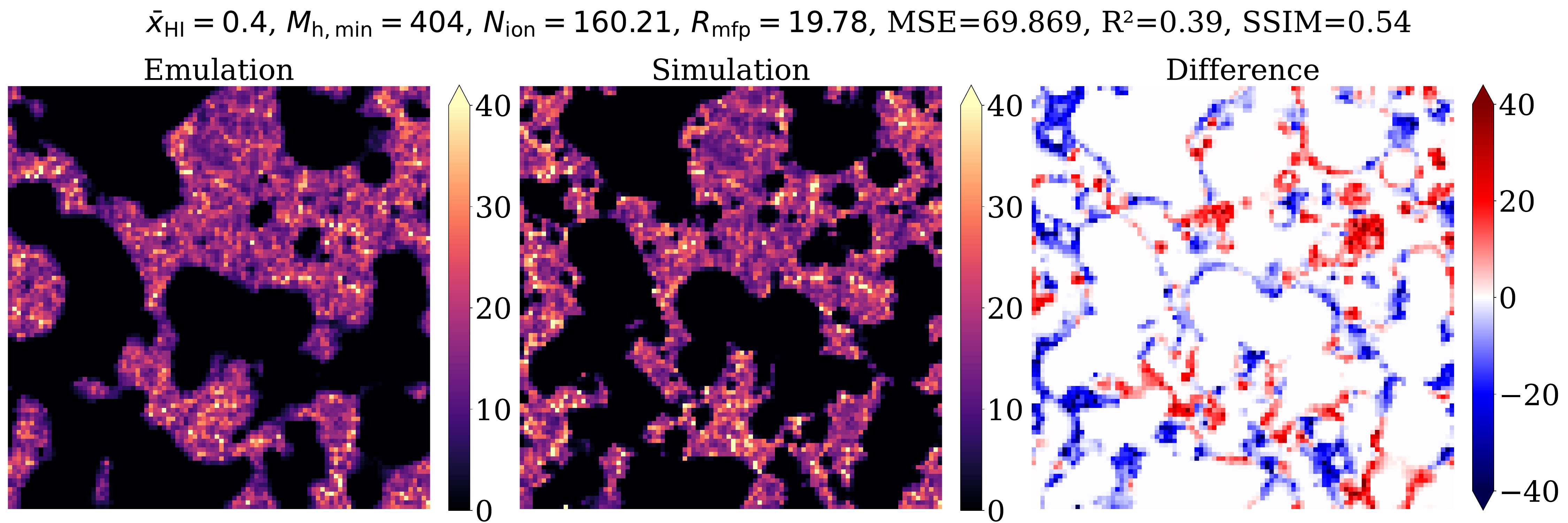}

    \caption{  Comparison between 21-cm brightness temperature fields (in mK) produced by emulation (\texttt{CosmoUiT}) and simulation (\texttt{ReionYuga}) for input fields generated using random seeds not used in training. The first two columns show the emulator and simulation outputs, while the third column gives the difference between them. Each title contains the mass-averaged neutral fraction, the EoR parameters (in the same units as the model inputs), and the corresponding metric scores.}
    \label{fig:prediction_unseen_21cm_rs}
\end{figure}

We also computed the 21-cm brightness temperature fields from the neutral fraction fields obtained from emulation and the simulation (Figure \ref{fig:prediction_unseen_21cm_rs}). In the seen input realizations scenario, the $\text {R}^2$ score remained almost unchanged while the SSIM improved. In contrast, for unseen input realizations, we observe a moderate reduction in the $\text{R}^2$ score (by $\approx 0.1-0.15$). This happens because $(\delta T_{21})$ depends on both the neutral fraction and the density field, so even small mismatches get amplified, and the overall variance in the brightness temperature field is higher. On the other hand, the SSIM values stay similar or improve slightly ($\approx 0.05 - 0.07$). This suggests that the model still captures the large-scale morphology and contrast of the 21-cm signal, even if the variance explained ($\text{R}^2$ score) is lower. 

\begin{figure}[htbp]
    \centering
    \includegraphics[width=0.55\linewidth]{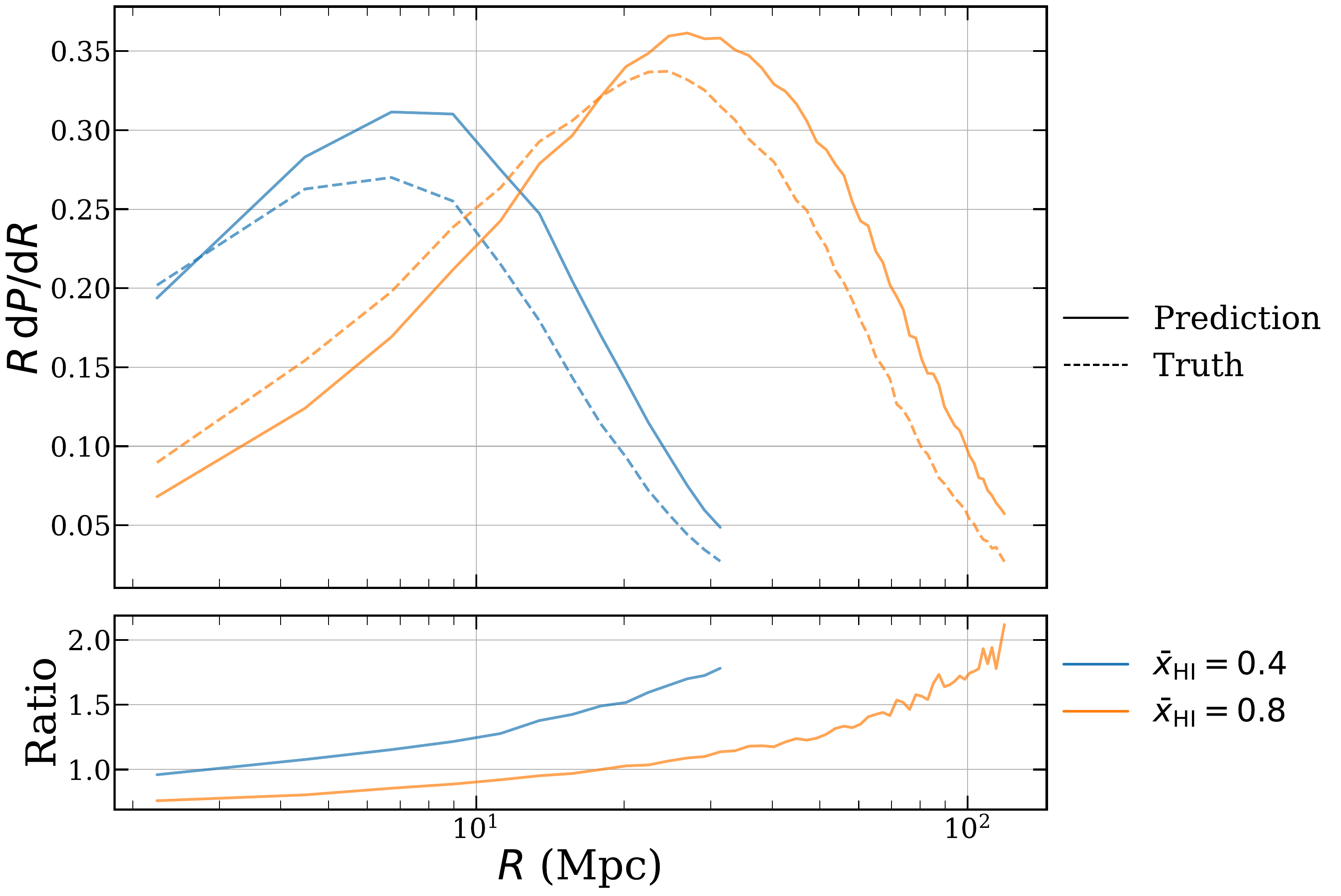}
    \caption{Bubble size distribution comparison for the two different neutral fractions of out-of-domain test data. The solid line represents the distribution obtained from the \texttt{CosmoUiT} prediction, and the dotted line represents the simulations. The blue and orange colours represent the mass-averaged neutral fractions of $0.4$ and $0.8$, respectively.}
    \label{fig:OOD_BSD}
\end{figure}

\begin{figure}[htbp]
    \centering
    \includegraphics[width=0.45\linewidth]{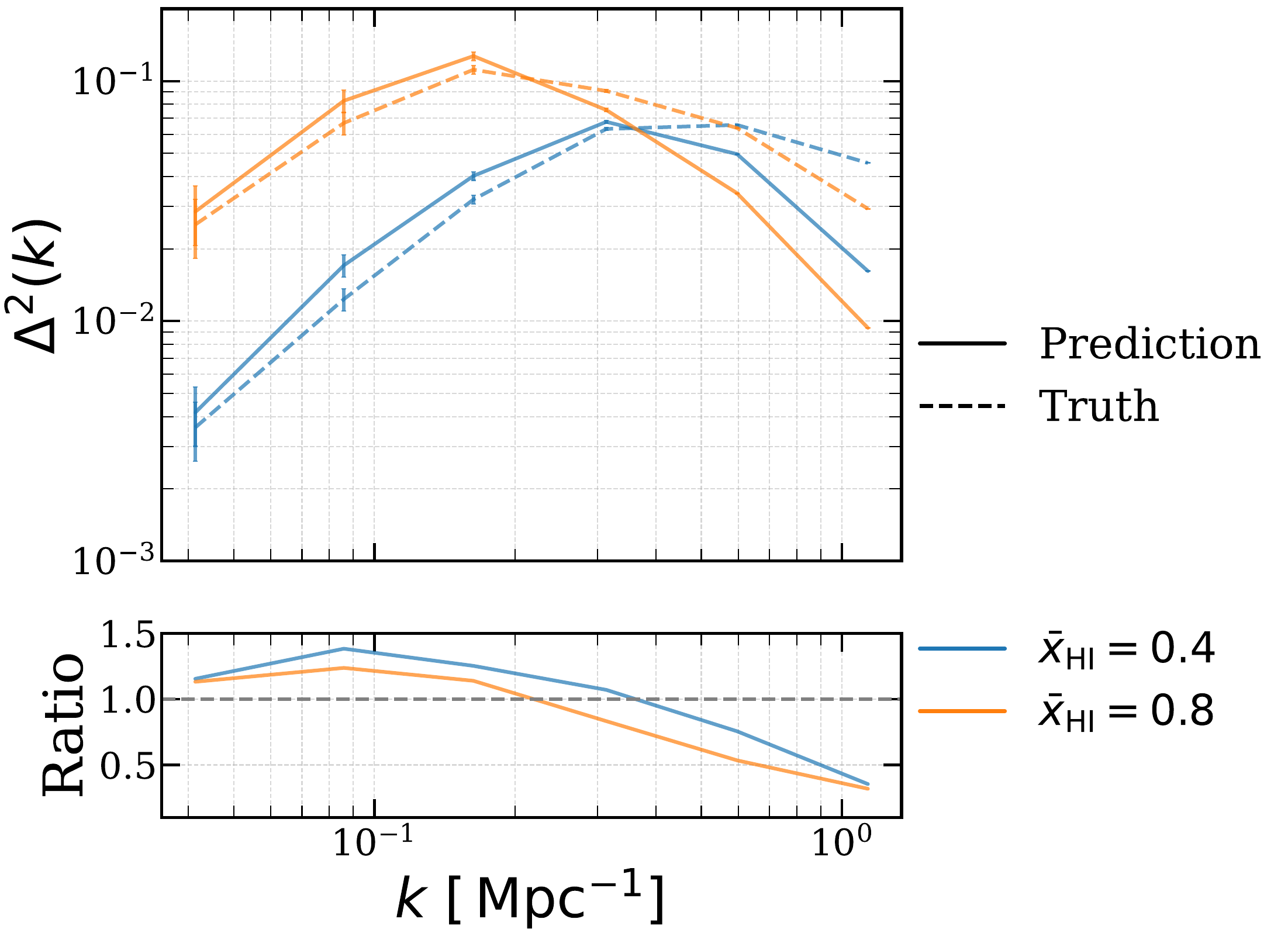}
    \includegraphics[width=0.45\linewidth]{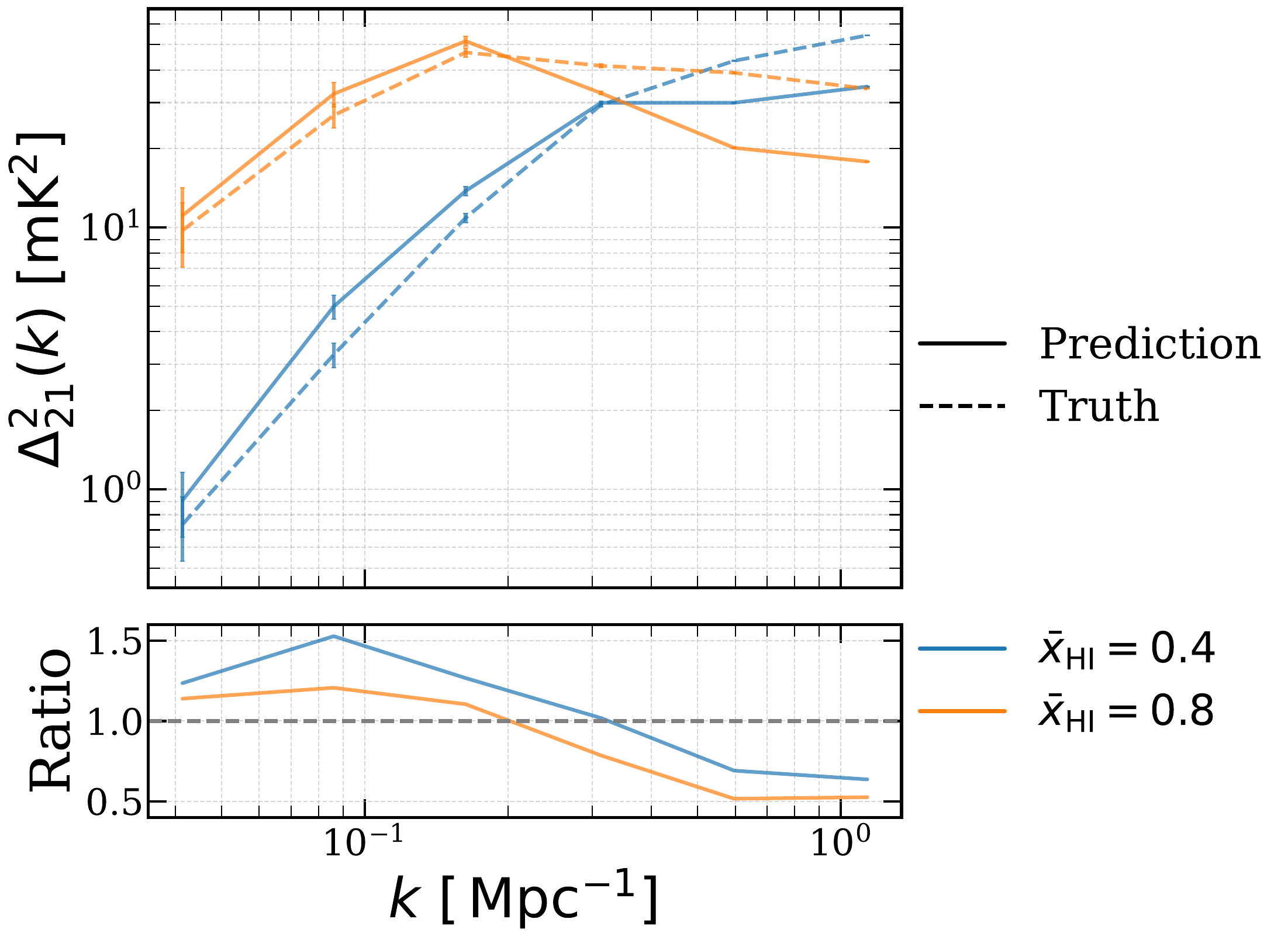}
    \caption{The left plot shows the comparison of $x_{HI}$ power spectrum, and the right one compares the 21-cm brightness temperature power spectra for the two different neutral fractions of out-of-domain test data. The solid line represents the distribution obtained from the \texttt{CosmoUiT} prediction, and the dotted line represents the simulations. The blue and orange colours represent the mass-averaged neutral fractions of $0.4$ and $0.8$, respectively. }
    \label{fig:OOD_PS}
\end{figure}

Furthermore, the comparison of the bubble distribution and power spectra is shown in Figures \ref{fig:OOD_BSD} and \ref{fig:OOD_PS}. The bubble size distribution of the predicted field (solid line) is overestimated for neutral fraction $0.4$ (shown in blue). Here, the overestimation of the distribution is also due to the fuzzy boundary problem. Since these test data are generated using a completely different initial random seed, the fuzziness at the boundaries of the ionized bubble between predicted and simulated field is very large, resulting in the large overestimation of bubble distribution compared to the test cases shown in Section \ref{sec:bubble_size_distribution}. For the neutral fraction $0.8$, it is initially underestimated and then overestimated. The initial underestimation happens because at this high neutral fraction, a large number of small-sized ionized bubbles, which are not captured by the \texttt{CosmoUiT}, as one can see in Figure \ref{fig:  prediction_unseen_rs}.

Similarly, in $x_{HI}$ and 21-cm power spectra (see Figure \ref{fig:OOD_PS}), the effect of the fuzzy boundaries and loss of small-scale features is evident as the power spectrum is considerably underestimated at large k-modes for these OOD test data compared to test data of the same initial seed (see Figures \ref{fig:xHI_power_spectrum} and \ref{fig:21cmBT_power_spectrum}). Additionally, we can see that at large scales it overestimates the power spectrum by a small factor. However, given that \texttt{CosmoUiT} can predict the fields with the power spectra within less than an order of magnitude for unseen initial seeds, it is very promising to move forward to develop a generalized model that can accurately predict the field for any random initial seed data.

The model produces parameter-specific outputs and captures the large-scale structures. These results confirm that the emulator has learned a generalized mapping from the input fields and astrophysical parameters to the corresponding ionization state, rather than memorizing specific spatial configurations from training. Further, it suggests that training this model on a few examples of input fields with varying initial random seeds would yield better results. This generalization ability is critical for applying the emulator to field-level inference scenarios, where the underlying initial conditions are inherently unknown and the cosmic variance needs to be taken into account.

\subsection{Parameter Inference using Power spectrum} \label{Parameter-inference}

We perform Bayesian parameter inference to estimate reionization model parameters, using the power spectrum as a summary statistic, to demonstrate the statistical robustness of the fields predicted by \texttt{CosmoUiT}. To forward model the signal, we use \texttt{CosmoUiT} to predict the 21-cm field from reionization parameters and estimate the power spectrum of the predicted field. The covariance in the likelihood is estimated as,
\begin{equation}
    \Sigma = \sigma_{\rm T}^2  = \sigma^2_{\rm SV} + \sigma^2_{\rm N},
\end{equation}
where $\sigma^2_{\rm SV}$ is the sample variance introduced due to the binning of the spherical power spectrum and $\sigma^2_{\rm N}$ is introduced by the system noise. system noise based on $1000$ hr of SKA-LOW observations. 
More details on the noise estimation and the choice of the instrumental parameters can be found in \cite{Mahida_2025}.
Ideally, this covariance should also consider the uncertainty associated with \texttt{CosmoUiT} prediction. However, our emulator is still deterministic and thus does not provide prediction uncertainty directly. We have tried to estimate the uncertainty of the emulator (see Appendix \ref{sec:uncertainty_qantification}); however, this is not a robust approach and requires further improvement. In our future work, we intend to use the Bayesian approach to estimate the emulator prediction uncertainty and then propagate it through the inference pipeline.  We have used the MH-MCMC sampler to sample the parameter space, which takes $ \sim 50,000 $ likelihood evaluations to constrain the parameters. This takes $\sim 1$ hr compared to a few hundred hrs if we had used the \texttt{ReionYuga} simulation.

\begin{figure}[htbp]
    \centering
    \includegraphics[width=0.45\linewidth]{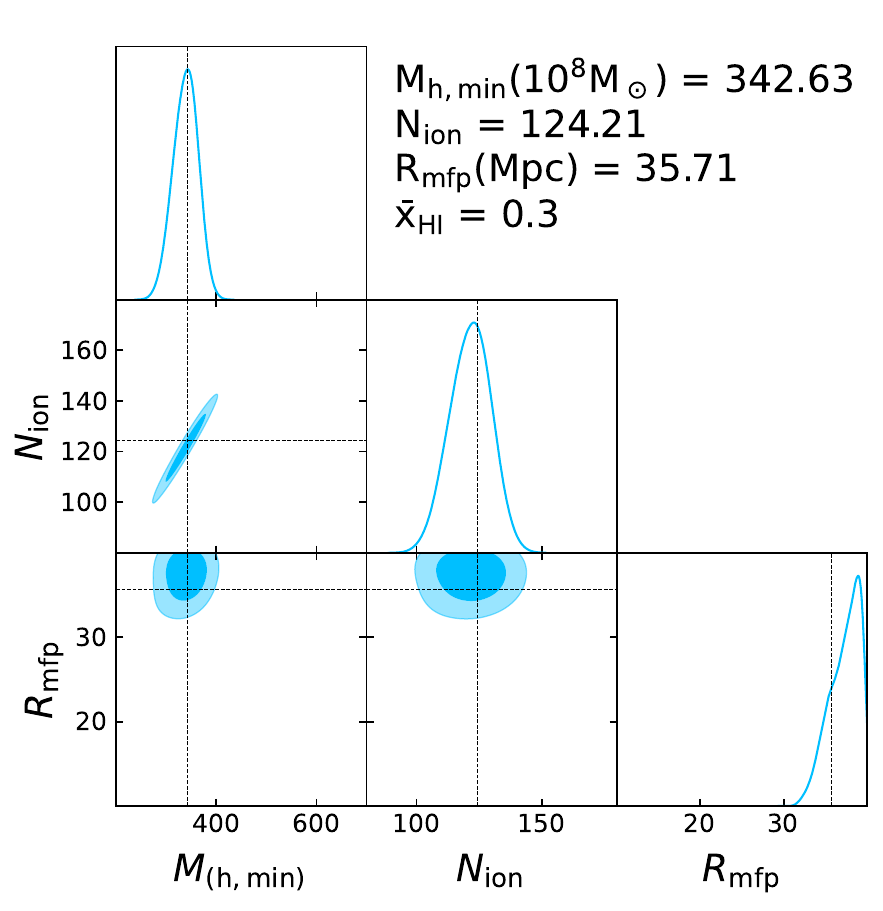}
    \includegraphics[width=0.45\linewidth]{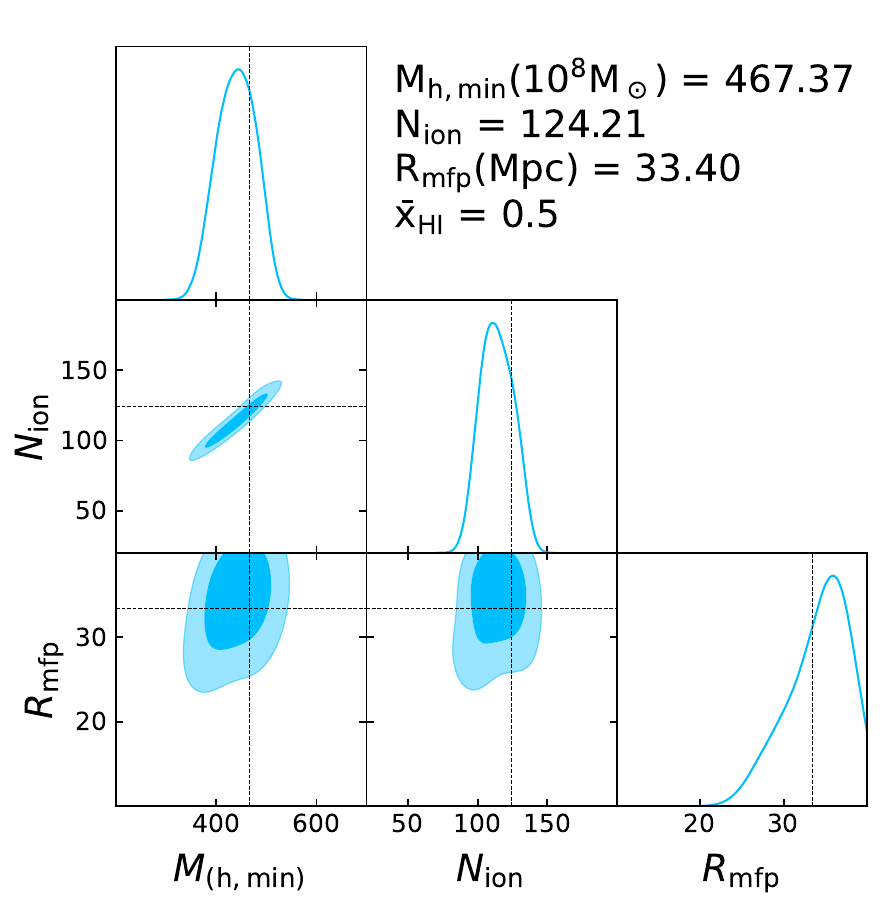}
    \includegraphics[width=0.45\linewidth]{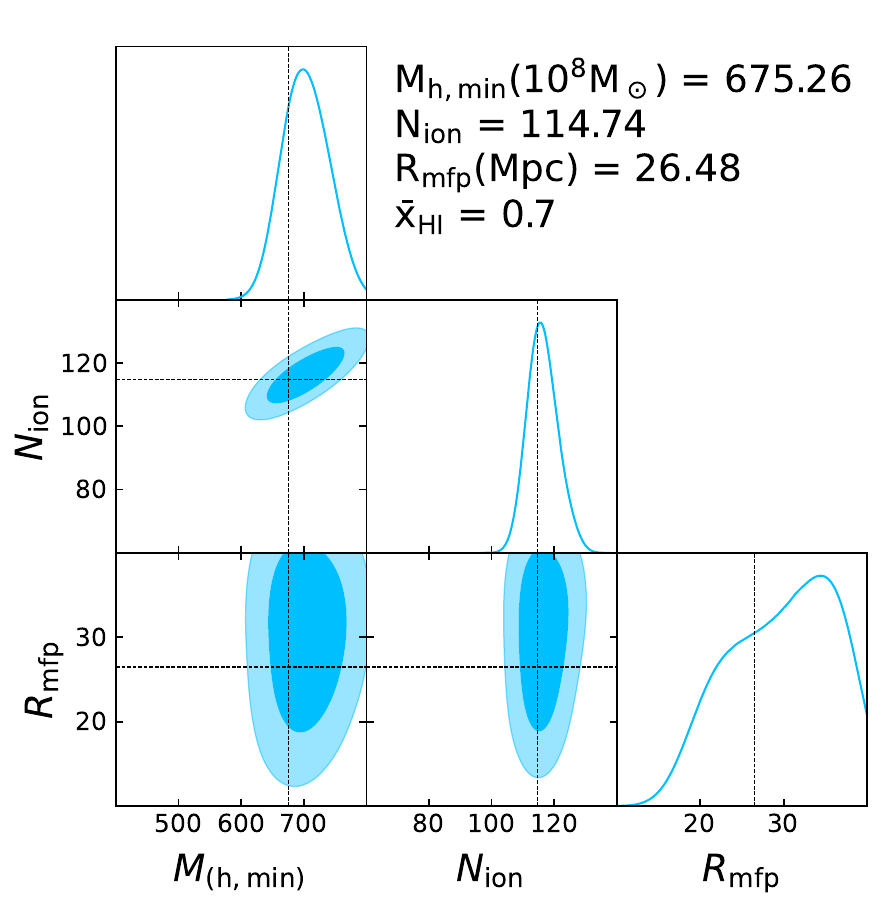}
    \caption{Marginalized posteriors of the reionization parameters for three different neutral fractions: $\bar{x} = 0.3$ (top left), $\bar{x} = 0.5$ (top right) and $\bar{x} = 0.7$ (bottom)}
    \label{fig: inference plot1}
\end{figure}

Constraint plots of the reionization parameters for three different reionization scenarios (neutral fractions: $0.3$, $0.5$ and $0.7$, see Figures \ref{fig:cosmouit96_predictions} and \ref{fig:prediction_21cm_all}) are shown in the Figure \ref{fig: inference plot1}. We have used the first $4$ $k$-bins of the 21-cm power spectra shown in Fig. \ref{fig:21cmBT_power_spectrum} as mock power spectra. Reionization parameters are considerably well constrained for all three neutral fractions. However, for the neutral fraction $0.7$, the mean free path $ \rm R_{mfp}$ shows degeneracy which is expected as for high values of $\rm M_{h, min}$ and $\rm N_{ion}$, mean free path has very little effect on the neutral fraction and overall morphology of signal resulting degenerate power spectra.

\section{Summary and Discussion}\label{sec:Summary}
The aim of this work has been to design a fast and accurate emulator for predicting 3D 21-cm brightness temperature fields during the Epoch of Reionization (EoR), using the underlying dark matter and halo density fields as inputs along with the three reionization parameters. We introduce \texttt{CosmoUiT}, a UNet integrated Vision Transformer architecture that can capture both the global morphology and local fluctuations of reionization while retaining sensitivity to the reionization parameters. This allows \texttt{CosmoUiT} to act as a field-level emulator, helping us bypass computationally expensive simulations and enabling parameter inference for next-generation 21-cm tomographic radio surveys.
Our main results are summarized below:
\begin{itemize}
    \item      Voxel-wise fidelity: \texttt{CosmoUiT} shows excellent agreement with reference simulations across a wide range of reionization parameters. Quantitative metrics (MSE, $R^2$, SSIM) consistently confirm that the model reproduces both global morphologies and local fluctuations with high accuracy. This highlights the emulator as a dependable framework for voxel-wise prediction of the 21-cm signal.
      
      \item Large and Small-Scale Morphologies: \texttt{CosmoUiT} is able to reconstruct the changing morphology of ionized regions across different reionization parameters. On large scales, the power spectra of both the ionization fields and 21-cm brightness temperature fields are reproduced with high accuracy, capturing the correct amplitude and slope. This indicates that the model is sensitive to global reionization topology. At small scales, the emulator exhibits mild power suppression. We observed that while predicting the ionization fields, it struggles to capture abrupt transitions at boundaries of the ionized regions and instead predicts more gradual variations, which causes a smoothing of sharp ionization boundaries. As a result, the emulator tends to predict slightly larger ionized regions than those inferred from the simulations.

      \item      Bubble-size distributions: In addition to power spectra, \texttt{CosmoUiT} closely captures the distribution of ionized bubble sizes throughout reionization parameter space. The agreement is strong across parameter space, with deviations arising mainly in cases that correspond to high ionization fractions. In these cases, boundary smoothing leads to a slight overestimation of large bubbles. 

      \item  Generalization to unseen initial conditions: A critical requirement for a deep learning model is its ability to generalize beyond the training domain. Our results demonstrate that \texttt{CosmoUiT} performs reliably well when tested on dark matter density and halo fields generated using an unseen initial random seed. This indicates that the model has learned a mapping from physical inputs and parameters to the ionization field, rather than memorizing specific spatial configurations. This is a crucial property for applications to inference pipelines.

    \item Comparison with previous approaches: Earlier CNN-based emulators, such as \texttt{PINION} and \texttt{CRADLE}, tackled different aspects of the emulation problem; however, they were restricted in scope. The \texttt{CRADLE} architecture captures large-scale features effectively through its slice-based framework, but struggles to reproduce small-scale structures due to the smoothing of the input fields. On the other hand, the \texttt{PINION} architecture achieves strong performance at small scales and incorporates physics-informed losses. The pixel-wise prediction and subsequent reconstruction of the 3D cubes limit the architecture's ability to accurately capture long-range dependencies. Moreover, both methods were trained on a fixed set of astrophysical parameters, which limits their applicability for parameter inference. The \texttt{CosmoUiT} addresses these limitations by using transformer encoders to capture global context and to embed parameters into the input fields so that the output becomes conditioned on three EoR parameters. 

    \item EoR simulation framework: The emulator is trained on ionization fields generated using the semi-numerical code \texttt{ReionYuga}, which models reionization using the excursion-set formalism by assuming that ionizing sources reside in dark matter halos above a minimum mass threshold, that the hydrogen density traces the underlying dark matter distribution, and that ionized regions grow according to an effective ionizing efficiency and a finite photon mean free path. While this framework efficiently captures the large-scale morphology of reionization and is well suited for generating large training datasets, extending the training to more sophisticated simulations that include additional physical processes, such as detailed radiative-transfer effects or more complex source prescriptions (e.g. \texttt{C$^2$-Ray} \cite{Mellema_2006, HirlingC2Ray}), represents a natural direction for future work.
    \item  Computational efficiency: Once trained, the \texttt{CosmoUiT} can generate a full 3D coeval $x_{\mathrm{HI}}$ or 21-cm brightness temperature cube for a given set of reionization parameters in $\sim 0.03$ seconds on an RTX A4000 (16 GB) GPU. Due to the inherently parallel nature of the model, multiple realizations can be generated simultaneously, depending on the available GPU memory. In comparison, a single semi-numerical reionization simulation typically requires $\sim 1$ hour for each parameter evaluation. Therefore, a Bayesian inference requiring $\sim 10^5$ likelihood evaluations would take $\mathcal{O}(10^5)$ hours using direct simulations, whereas the same analysis can be performed in $\mathcal{O}(10^3)$ seconds using \texttt{CosmoUiT}. This significant reduction in computational cost makes field-level Bayesian inference over the EoR parameter space computationally tractable.
    
    \item Possible improvements: The \texttt{CosmoUiT} demonstrates excellent performance in terms of accuracy and speed; however, there is a significant amount of improvement that we plan to achieve in future. Presently, it emulates the coeval cubes at a single redshift snapshot (z=$7$). We will extend it to emulate entire lightcone cuboids that capture the redshift evolution of the 21-cm signal as expected in the surveys with SKA-Low.
    Additionally, we are further developing this emulator to emulate the 21-cm field with higher spatial resolution (i.e., without downsampling the field due to GPU memory constraints), including the effect of the redshift space distortions and, most importantly, to get a proper understanding and quantification of the emulator uncertainty. Ignoring them in the likelihood function will lead to biased parameter estimates. Although the current approach shows a strong correlation between uncertainty and error, we plan to improve it using Bayesian neural networks. In addition to the astrophysics of the IGM and cosmology, the 21-cm signal is also affected by residual foregrounds, system noise, and telescope effects. These effects are not added to the current training data. To produce realistic data, the emulator should be trained to incorporate these effects and accurately reproduce them. Additionally, we have trained our model on input fields generated using fixed initial random seeds. Although it generalises reasonably well in cases of input fields obtained using unseen (during the training) initial random seeds, explicitly training on such variations would improve the prediction accuracy. This would also allow cosmic variance to be properly accounted for when generating training ensembles. Once trained and validated with the corrections outlined above, the emulator can be applied to field-level inference of astrophysical parameters using mock observations. This emulator will serve as a powerful tool for interpreting the future 3D tomographic observations performed with the SKA.
    
\end{itemize}

\acknowledgments
YM acknowledges the financial support from the Department of Science and Technology, Government of India, through the INSPIRE Fellowship. SM thanks the Science and Engineering Research Board (SERB) and the Department of Science and Technology (DST), Government of India, for financial support through Core Research Grant No. CRG/2021/004025 titled “Observing the Cosmic Dawn in Multicolor using Next Generation Telescopes”. LN acknowledges the financial support from the Department of Science and Technology, Government of India, through the INSPIRE Fellowship.

\appendix
\section{Architectural Strategies}\label{sec:architectural_strategies}
This section provides a detailed discussion of the architectural strategies developed before \texttt{CosmoUiT}, along with their model summaries, corresponding results, and metric scores. We initially experimented with CosmoViT and CosmoUNet, training and evaluating them alongside CosmoUiT on input–output pairs of resolution $48^3$. For consistency, all models were trained for $100$ epochs with a batch size of $16$ on an NVIDIA RTX A4000 GPU with 16GB of memory. Core architectural components, such as patch size, activation functions, embedding dimensions, the number of feature maps, and the number of attention heads (if any), were kept fixed across all variants. This setup enabled a fair comparison without excessive resource consumption, though training time varied with model complexity. Based on validation metrics, CosmoUiT demonstrated the most consistent results across different parameter combinations. Insights from these comparisons guided the design of the final architecture, where the resolution of the fields was increased from $48^3$ to $96^3$, referred to as CosmoUiT.

\subsection{\texttt{CosmoViT}}
\begin{figure}[htbp]
    \centering
    \includegraphics[width=0.98\textwidth]{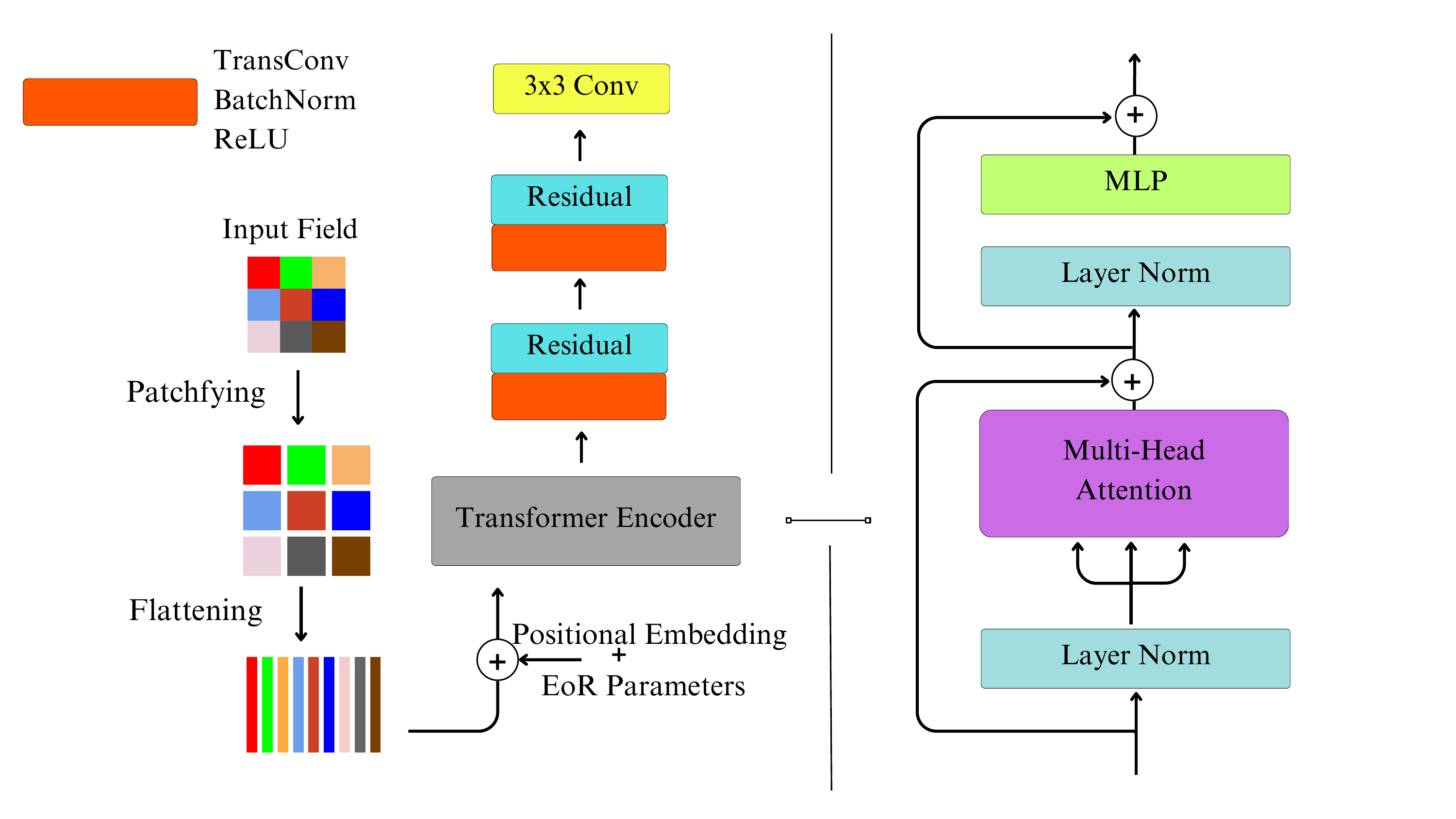} 
    \caption{Model Architecture of CosmoViT.}
    \label{fig:CosmoViTArchitecture}
\end{figure}

The \texttt{CosmoViT} architecture is adapted from the vision transformer-based segmentation model proposed in \cite{Gunduc_2021}, originally designed for binary image segmentation tasks. Given the binary nature of the neutral hydrogen fraction fields (with values close to $0$ in ionized regions and $1$ in neutral regions), this architecture was considered suitable for emulating reionization fields. We modified it to take 3D fields and $3$ reionization parameters as inputs. As illustrated in Figure \ref{fig:CosmoViTArchitecture}, the model begins by dividing the input volumes into non-overlapping $8^3$ patches, which are flattened and linearly projected into $128$-dimensional tokens. Learnable positional embeddings are added, and the EoR parameters are encoded and appended as additional tokens. The token sequence is passed through a Transformer encoder comprising $4$ layers, each with $8$ attention heads. The output tokens are reshaped into cubes and upsampled using a series of transpose convolutions and residual connections to reconstruct the output field at the original resolution. A summary of the architectural parameters is provided in Table \ref{tab:cosmovit_summary}.

\begin{figure}[htbp]
    \centering
    \includegraphics[width=0.98\textwidth]{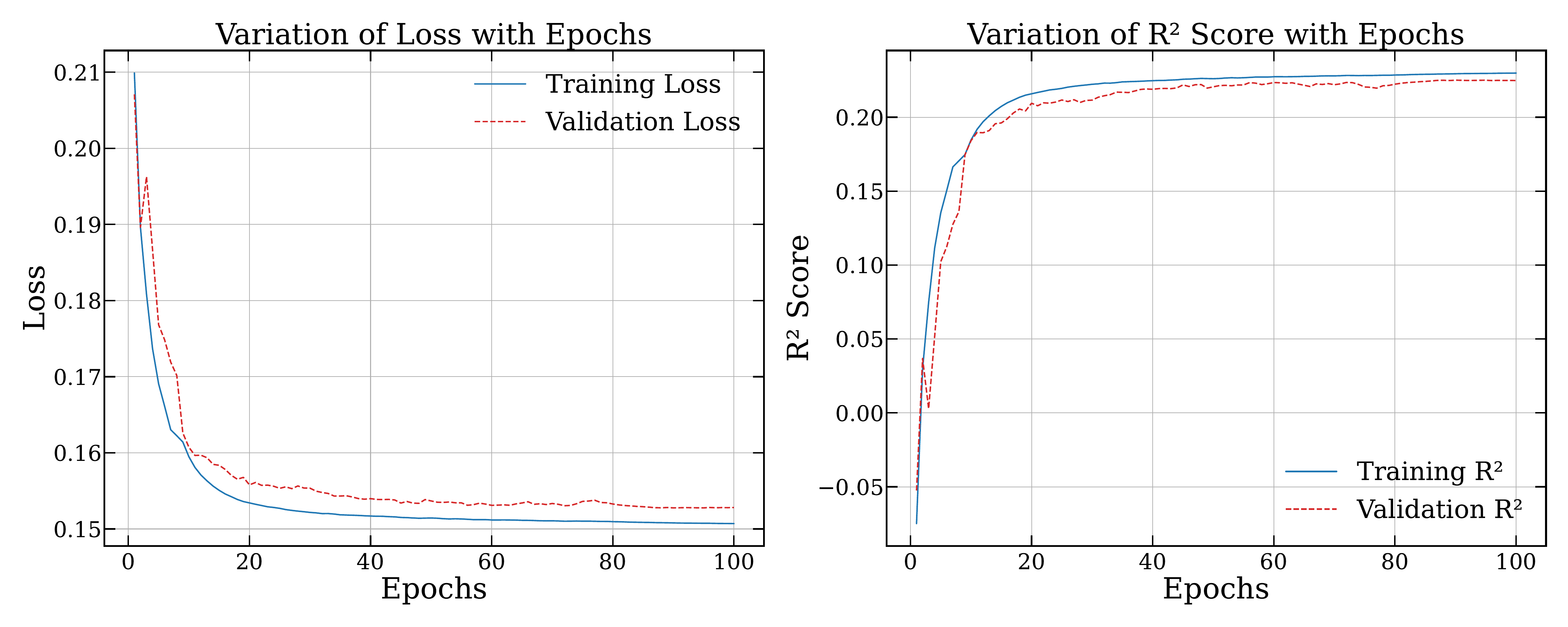}
    \caption{Variation in training and validation loss over the training iterations.}  
    \label{fig:CosmoViTbehavior}
\end{figure}

\begin{figure}[htbp]
    \centering
    
    \includegraphics[width=0.97\textwidth]{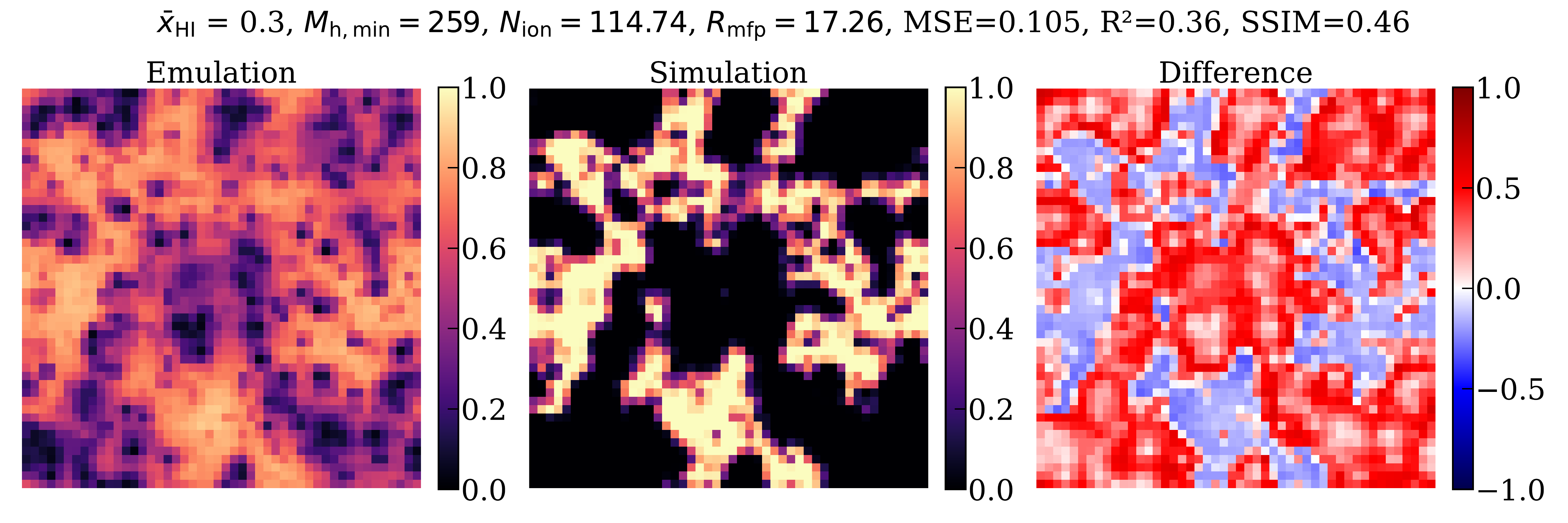}
    \vspace{0.5em}
    
    \includegraphics[width=0.97\textwidth]{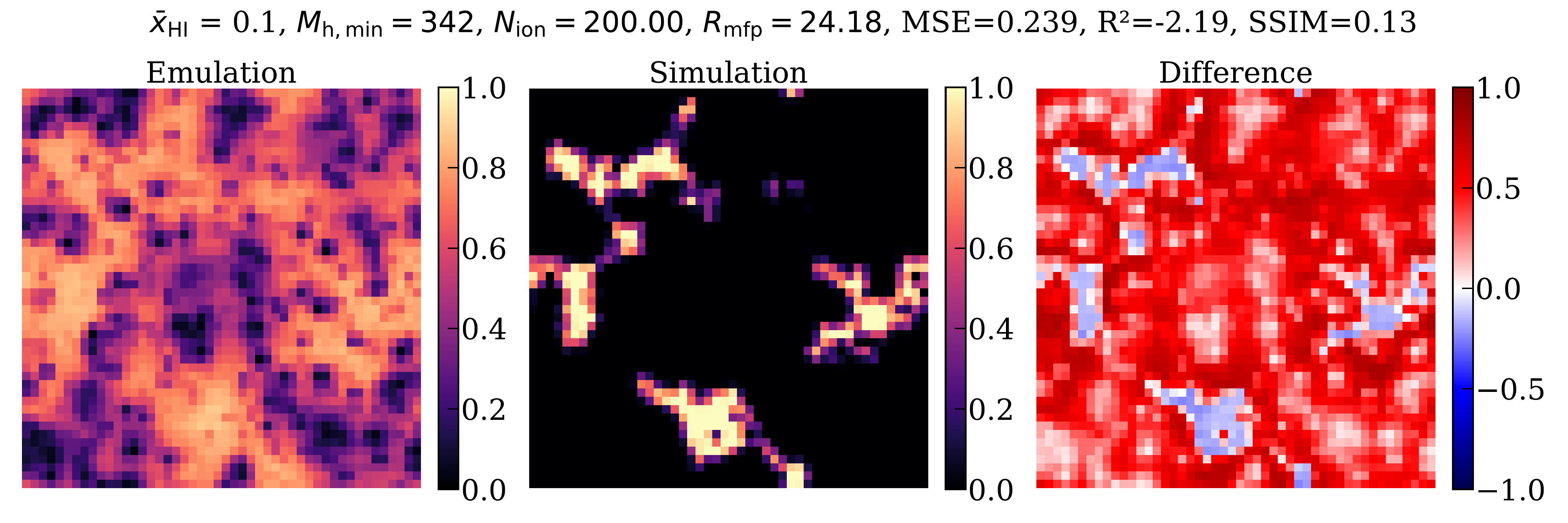}
    \caption{Comparison between $x_{\mathrm{HI}}$ fields produced by emulation (CosmoViT) and simulation. In the first two columns, $1$ corresponds to neutral regions and $0$ corresponds to ionized regions. The third column gives the difference between these two fields. Each title contains the mass-average neutral fraction, the EoR parameter values in the units used to feed them to the architecture, and the metric scores.}
   \label{fig:CosmoViT100}
\end{figure}

Figure \ref{fig:CosmoViTbehavior} presents the training and validation loss curves for the \texttt{CosmoViT} model across training iterations. While both losses decrease exponentially, they converge to relatively high values of $0.151$ (training) and $0.153$ (validation), indicating limited learning. Correspondingly, the R$^2$ scores remain low, at $0.23$ for training and $0.22$ for validation. It suggests that the model struggles to capture the variability in the output data. This limitation is visually evident in Figure \ref{fig:CosmoViT100}, where the model fails to generate outputs that reflect different reionization parameter combinations. Instead, it produces nearly identical, parameter-agnostic fields. This behavior can be attributed to the model’s original design for image translation tasks, which typically assumes a one-to-one mapping between input and output. In our case, the input fields remain fixed while the variation arises primarily from the three reionization parameters, leading to a complexity that this architecture is not well suited to handle.

\begin{table}[htbp]
\centering
\begin{tabular}{|l|c|c|l|}
\hline
\textbf{Component} & \textbf{No. of Feature Maps} & \textbf{Filter Size} & \textbf{Activation Function} \\ \hline
\textbf{Patchifying \& Embedding}       &                    &                     &                             \\ \hline
Patch Embedding       & $216$ tokens                    & $8 \times 8 \times 8$                    & -                            \\ \hline
Projection Layer & $512 \xrightarrow{} 128$                  & -                         & -                            \\ \hline
Position Embedding &       $(1, 216, 128)$          & -                         & -                            \\ \hline
Parameter Embedding    & $3 \xrightarrow{} 128 $                    & -                         & -                            \\ \hline
\textbf{Transformer Encoder} & 4 Layers               & -                         & ReLU                         \\ \hline
\hspace{0.5cm} - Self-Attention  & 8 Heads               & -                         & -                            \\ 

 & $(n, 8, 2 \times (216 + 3), 16) $              &                          &                             \\ \hline

\hspace{0.5cm} - Feedforward     & $128 \xrightarrow{} 256 \xrightarrow{} 128 $             & -                         & ReLU                         \\ \hline
\hspace{0.5cm} - Layer Normalization & 128                  & -                         & -                            \\ \hline

\textbf{Upsampler} &  &  &  \\ \hline
ConvTranspose 1 & 256 & $3 \times3\times3$ & Leaky ReLU \\ \hline
Residual Block 1 & 256 & $3 \times3\times3$ & ReLU \\ \hline
ConvTranspose 2 & 128 & $3 \times3\times3$ & Leaky ReLU \\ \hline
Residual Block 2 & 128 & $3 \times3\times3$ & ReLU \\ \hline
ConvTranspose 3 & 64 & $3 \times3\times3$ & Leaky ReLU \\ \hline
Residual Block 3 & 64 & $3 \times3\times3$ & ReLU \\ \hline
ConvTranspose 4 & 32 & $9 \times9\times9$ & Leaky ReLU \\ \hline
Residual Block 4 & 32 & $3 \times3\times3$ & ReLU \\ \hline
ConvTranspose 5 & 16 & $9 \times9\times9$ & Leaky ReLU \\ \hline
Residual Block 5 & 16 & $3 \times3\times3$ & ReLU \\ \hline
Final Conv & 1 & $3 \times3\times3$ & ReLU \\ \hline
\end{tabular}
\caption{Summary of the CosmoViT (base model) architecture.}
\label{tab:cosmovit_summary} 
\end{table}

\subsection{\texttt{CosmoUNet}}
\texttt{CosmoUNet} follows a standard UNet architecture \cite{ROnneberger_2015}, adapted to process two 3D input fields and the three reionization parameters. Unlike \texttt{CosmoUiT}, which incorporates a Transformer encoder and integration of parameters at two stages, \texttt{CosmoUNet} takes the dark matter and halo fields as direct inputs and integrates the three reionization parameters only at the bottleneck layer. The encoder and decoder configurations mirror those described in Section \ref{sec:uit_description}. The key difference lies in the absence of the transformer encoder and the use of static feature encoding, meaning that the feature maps extracted in the encoder remain unchanged across different combinations of reionization parameters. A schematic of the architecture is shown in Figure \ref{fig:CosmoUNetArchitecture}, and its key configurations are summarized in Table \ref{table:cosmounet_summary}.

\begin{table}[htbp]
\centering
\begin{tabular}{|l|c|c|c|}
\hline
\textbf{Component} & \textbf{No. of Feature Maps} & \textbf{Filter Size} & \textbf{Activation Function} \\ \hline
\textbf{Encoder}          &                       &                           &                              \\ \hline
DoubleConv Layer (enc1)         & 32         & $3\times3\times3$                     & ReLU                         \\ \hline
DoubleConv Layer (enc2)         & 32         & $3\times3\times3$                     & ReLU                         \\ \hline
DoubleConv Layer (enc3)         & 64        & $3\times3\times3$                     & ReLU                         \\ \hline
DoubleConv Layer (enc4)         & 128        & $3\times3\times3$                     & ReLU                         \\ \hline
\textbf{Bottleneck}             &  $128+3 $                     &                           &                              \\ \hline
DoubleConv (with Parameters)    & $131 \xrightarrow{} 512$        & $3\times3\times3$                     & ReLU                         \\ \hline
\textbf{Decoder}          &                       &                           &                              \\ \hline
ConvTranspose3D (upconv4)       & 256        & $2\times2\times2$                     & -                            \\ \hline
DoubleConv (dec4)               & 256        & $3\times3\times3$                     & ReLU                         \\ \hline
ConvTranspose3D (upconv3)       & 128        & $2\times2\times2$                     & -                            \\ \hline
DoubleConv (dec3)               & 128        & $3\times3\times3$                     & ReLU                         \\ \hline
ConvTranspose3D (upconv2)       & 64         & $2\times2\times2$                     & -                            \\ \hline
DoubleConv (dec2)               & 64         & $3\times3\times3$                     & ReLU                         \\ \hline
ConvTranspose3D (upconv1)       & 32         & $2\times2\times2$                     & -                            \\ \hline
DoubleConv (dec1)               & 32         & $3\times3\times3$                     & ReLU                         \\ \hline
Final Conv Layer                & 1             & $3\times3\times3$                     & -                            \\ \hline
\end{tabular}
\caption{Summary of the CosmoUNet (base model) architecture.}
\label{table:cosmounet_summary}
\end{table}

\begin{figure}[htbp]
    \centering
    \includegraphics[width=0.98\textwidth]{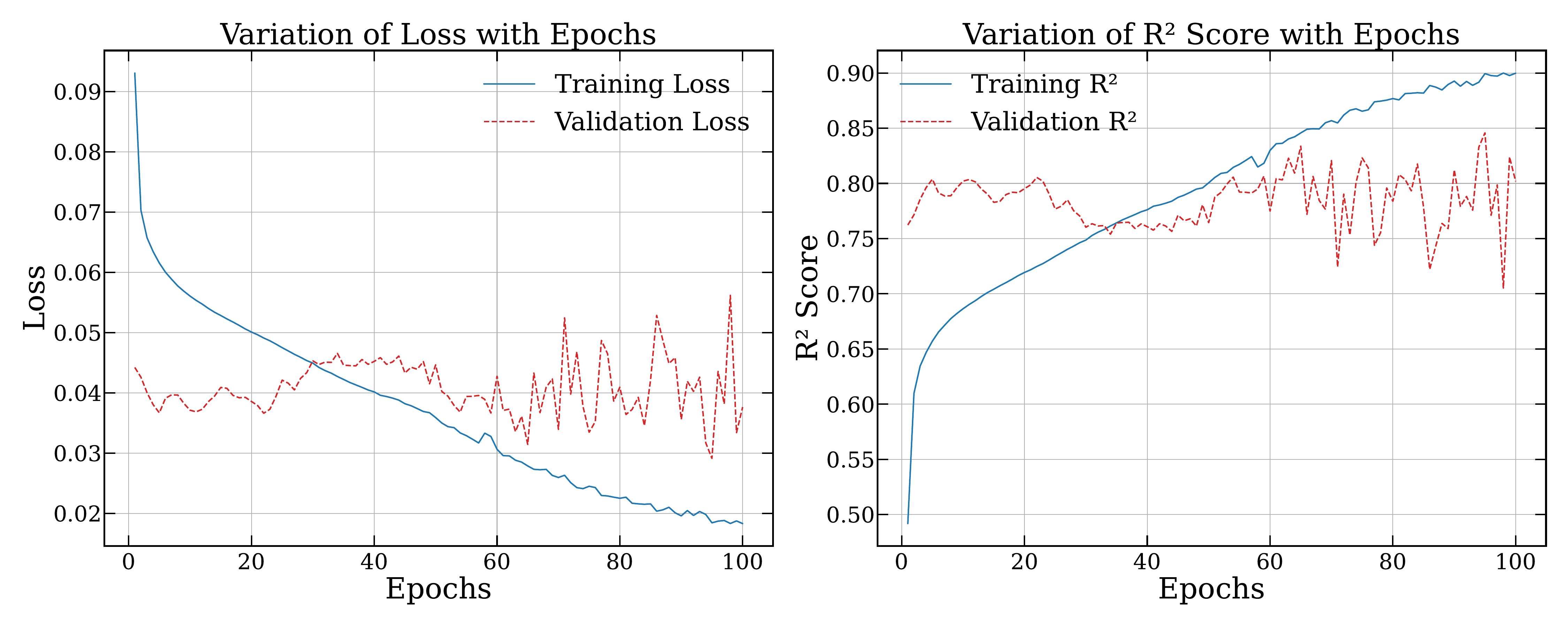}  
    \caption{The plot shows variations of MSE and $\text{R}^2$ for CosmoUNet for training and validation data over epochs.}
    \label{fig:CosmoUNetbehavior}
\end{figure}

\begin{figure}[htbp]
    \centering
    
    \includegraphics[width=0.97\textwidth]{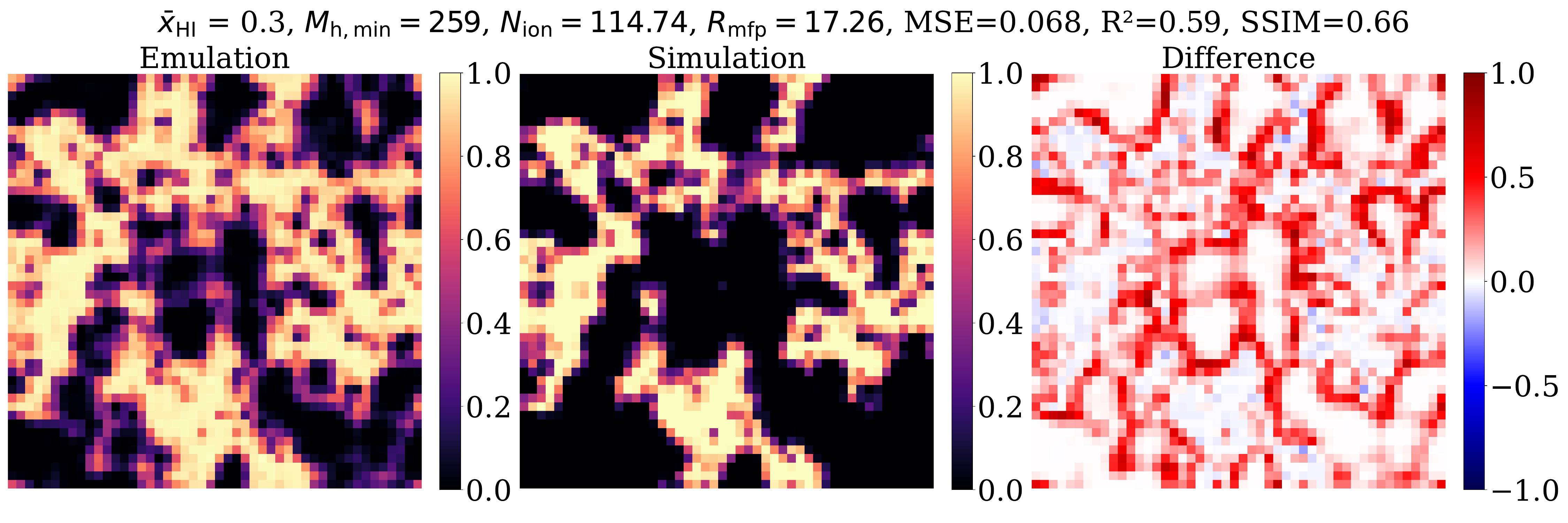}
    \vspace{0.5em}
    
    \includegraphics[width=0.97\textwidth]{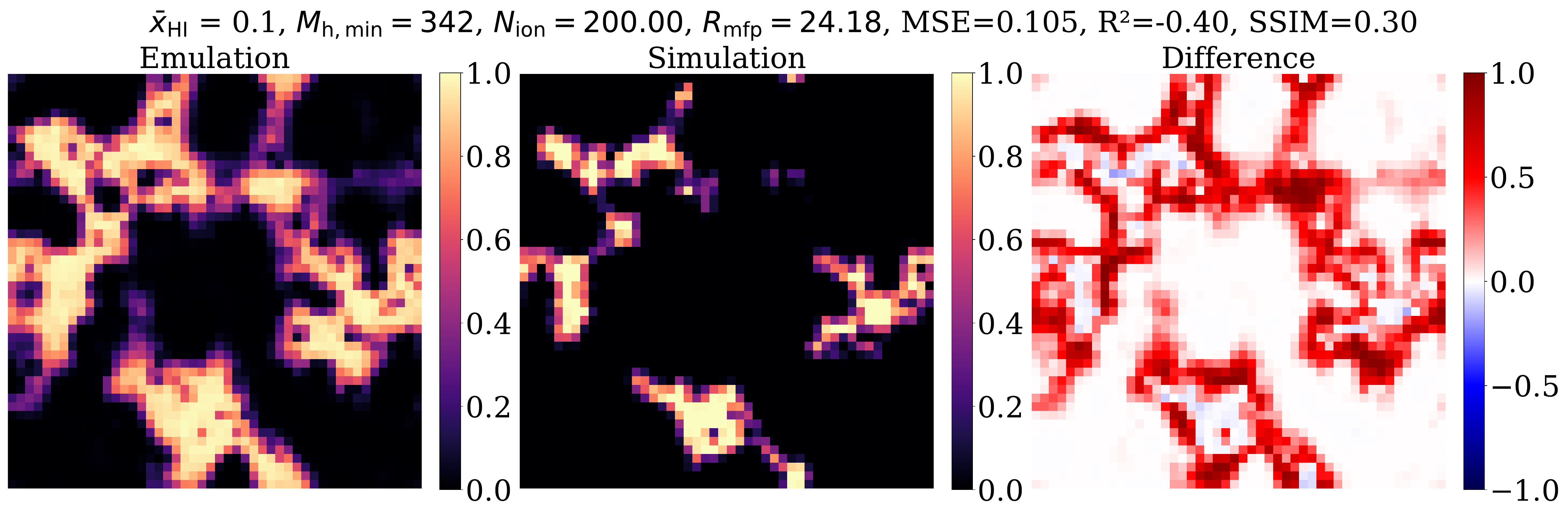}
    \caption{Comparison between \(x_{HI}\) fields produced by emulation (CosmoUNet) and simulation (ReionYuga).}
   \label{fig:CosmoUNet100}
\end{figure}

As shown in Figure \ref{fig:CosmoUNetbehavior}, while the training loss steadily decreases over the epochs, the validation loss remains relatively flat and begins to fluctuate once it is surpassed by the training loss. This indicates that the model fits well with the training data but fails to generalise to unseen data. This is an example of overfitting, where the model memorises the training examples rather than learning the underlying patterns. This limitation is reflected in the predicted output fields displayed in Figure \ref{fig:CosmoUNet100}. Although the outputs appear visually distinct across different combinations of reionization parameters, the mean squared error (MSE) remains high. The model consistently overestimates the extent of neutral regions, resulting in positive MSE values shown by red regions.

The reason behind this is the static feature encoding. Since UNet is primarily designed for image-to-image translation tasks involving one-to-one mappings, it relies heavily on variations in the input to produce corresponding changes in the output. In this case, the dark matter and halo fields remain fixed for all combinations of reionization parameters, and the parameters are introduced only at the bottleneck stage of the network. This late integration limits their ability to influence the model’s predictions. As a result, the network struggles to capture the variability driven by different parameter values, leading to poor parameter-specific generalization. This becomes more evident when we make predictions across all the parameter combinations available to us (see Section \ref{sec:copmarison_between_models}).

\subsection{\texttt{CosmoUiT48}}
This architecture was developed after identifying the limitations of \texttt{CosmoViT} and \texttt{CosmoUNet}, and it served as a precursor to the high-resolution \texttt{CosmoUiT} model described in Section \ref{sec:uit_description}. The key differences are lower input and output resolutions, fewer tokens, and fewer transformer encoder layers. This version uses four encoder layers, each with four attention heads, similar to the configuration used in \texttt{CosmoViT}. Additionally, the depth of the UNet architecture is reduced because the lower spatial resolution of the input fields limits the number of valid downsampling operations that can be performed without excessive loss of spatial information. Introducing a transformer encoder before the UNet, as done in \texttt{CosmoUiT}, addresses this issue by embedding the parameter information into the feature representation early in the network, thereby enabling parameter-specific predictions.

\begin{table}[htbp]
\centering
\begin{tabular}{|l|c|c|c|}
\hline
\textbf{Component} & \textbf{No. of Feature Maps} & \textbf{Filter Size} & \textbf{Activation Function} \\ \hline
\textbf{Vision Transformer}       &                    &                     &                             \\ \hline
Patch Embedding       & $216$ tokens                    & $8 \times 8 \times 8$                    & -                            \\ \hline
Projection Layer & $512 \xrightarrow{} 128$                  & -                         & -                            \\ \hline
Position Embedding &       $(1, 216, 128)$          & -                         & -                            \\ \hline
Parameter Embedding    & $3 \xrightarrow{} 128 $                    & -                         & -                            \\ \hline

\textbf{Transformer Encoder} & 4 Layers               & -                         & ReLU                         \\ \hline
\hspace{0.5cm} - Self-Attention  & 8 Heads               & -                         & -                            \\ 

 & (n, 8, 216+3, 16)               &                          &                             \\ \hline

\hspace{0.5cm} - Feedforward     & $128 \xrightarrow{} 256 \xrightarrow{} 128 $             & -                         & ReLU                         \\ \hline
\hspace{0.5cm} - Layer Normalization & 128                  & -                         & -                            \\ \hline
\textbf{UNet3D}          &          (Reconstructed Output)             &                           &                              \\ \hline
\textbf{Encoder}          &                       &                           &                              \\ \hline
DoubleConv Layer (enc1)         & 32         & $3\times3\times3$                     & ReLU                         \\ \hline
DoubleConv Layer (enc2)         & 32         & $3\times3\times3$                     & ReLU                         \\ \hline
DoubleConv Layer (enc3)         & 64        & $3\times3\times3$                     & ReLU                         \\ \hline
DoubleConv Layer (enc4)         & 128        & $3\times3\times3$                     & ReLU                         \\ \hline
\textbf{Bottleneck}             &  $128+3 $                     &                           &                              \\ \hline
DoubleConv (with Parameters)    & $131 \xrightarrow{} 512$        & $3\times3\times3$                     & ReLU                         \\ \hline
\textbf{Decoder}          &                       &                           &                              \\ \hline
ConvTranspose3D (upconv4)       & 256        & $2\times2\times2$                     & -                            \\ \hline
DoubleConv (dec4)               & 256        & $3\times3\times3$                     & ReLU                         \\ \hline
ConvTranspose3D (upconv3)       & 128        & $2\times2\times2$                     & -                            \\ \hline
DoubleConv (dec3)               & 128        & $3\times3\times3$                     & ReLU                         \\ \hline
ConvTranspose3D (upconv2)       & 64         & $2\times2\times2$                     & -                            \\ \hline
DoubleConv (dec2)               & 64         & $3\times3\times3$                     & ReLU                         \\ \hline
ConvTranspose3D (upconv1)       & 32         & $2\times2\times2$                     & -                            \\ \hline
DoubleConv (dec1)               & 32         & $3\times3\times3$                     & ReLU                         \\ \hline
Final Conv Layer                & 1             & $3\times3\times3$                     & -                            \\ \hline
\end{tabular}
\caption{Summary of the CosmoUiT (base model) architecture.}
\label{table:cosmouit_summary}
\end{table}

\begin{figure}[htbp]
    \centering
    \includegraphics[width=0.98\textwidth]{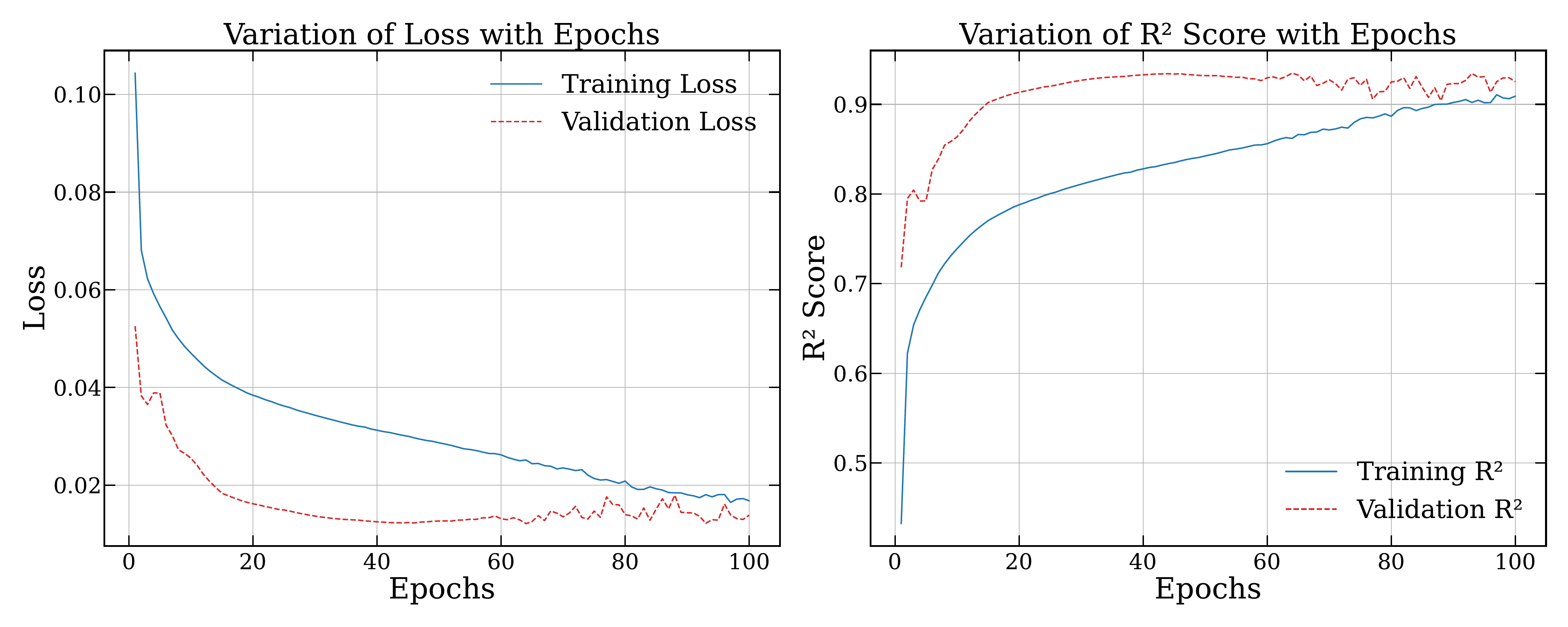}  
    \caption{The plot shows variations of MSE and $\text{R}^2$ for \texttt{CosmoUiT48} for training and validation data over epochs.}
    \label{fig:CosmoUiTbehavior}
\end{figure}

\begin{figure}[htbp]
    \centering
    
    \includegraphics[width=0.97\textwidth]{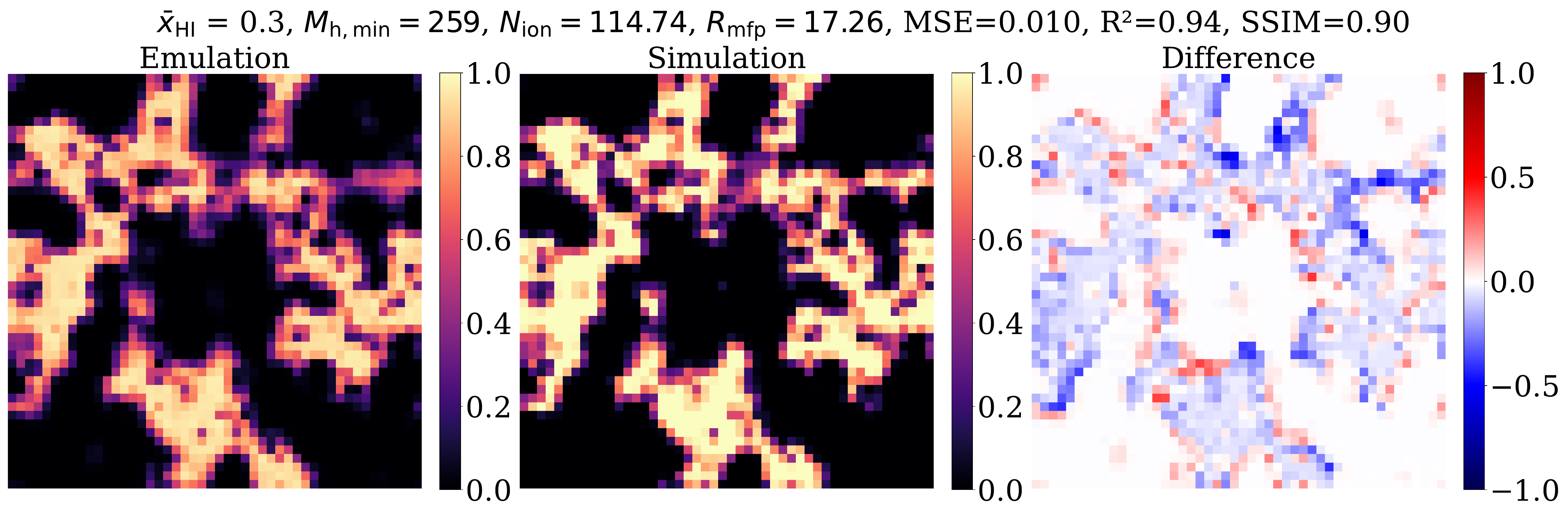}
    \vspace{0.5em}
    
    \includegraphics[width=0.97\textwidth]{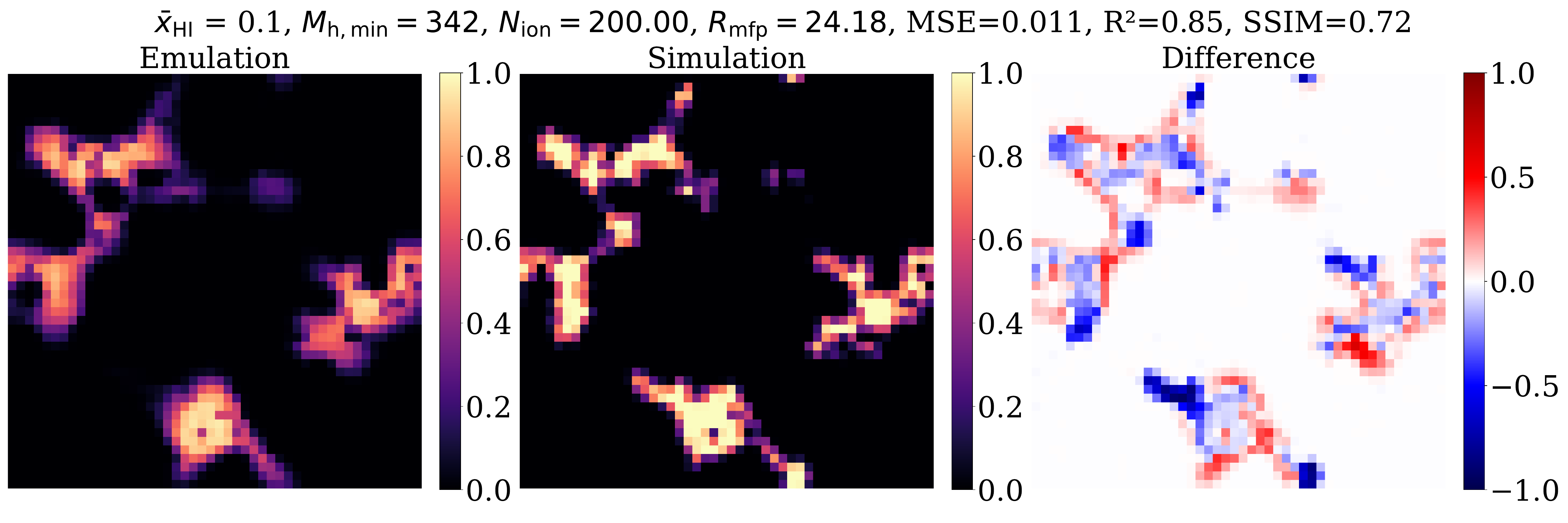}
    \caption{Comparison between \(x_{HI}\) fields produced by emulation (\texttt{CosmoUiT48}) and simulation.}
   \label{fig:CosmoUiT100}
\end{figure}

Figure \ref{fig:CosmoUiTbehavior} illustrates the variation of the mean squared error (MSE) loss and the $\text{R}^2$ score over training epochs. The validation loss shows a rapid initial decline and then stabilizes with minor fluctuations, while the training loss continues to decrease gradually. The corresponding predictions, shown in Figure \ref{fig:CosmoUiT100}, demonstrate that the model successfully generates parameter-specific outputs. Similar behavior is consistently observed across the full range of reionization parameter combinations.

\subsection{Comparison} \label{sec:copmarison_between_models}
\begin{figure}
    \centering
    \includegraphics[width=0.98\linewidth]{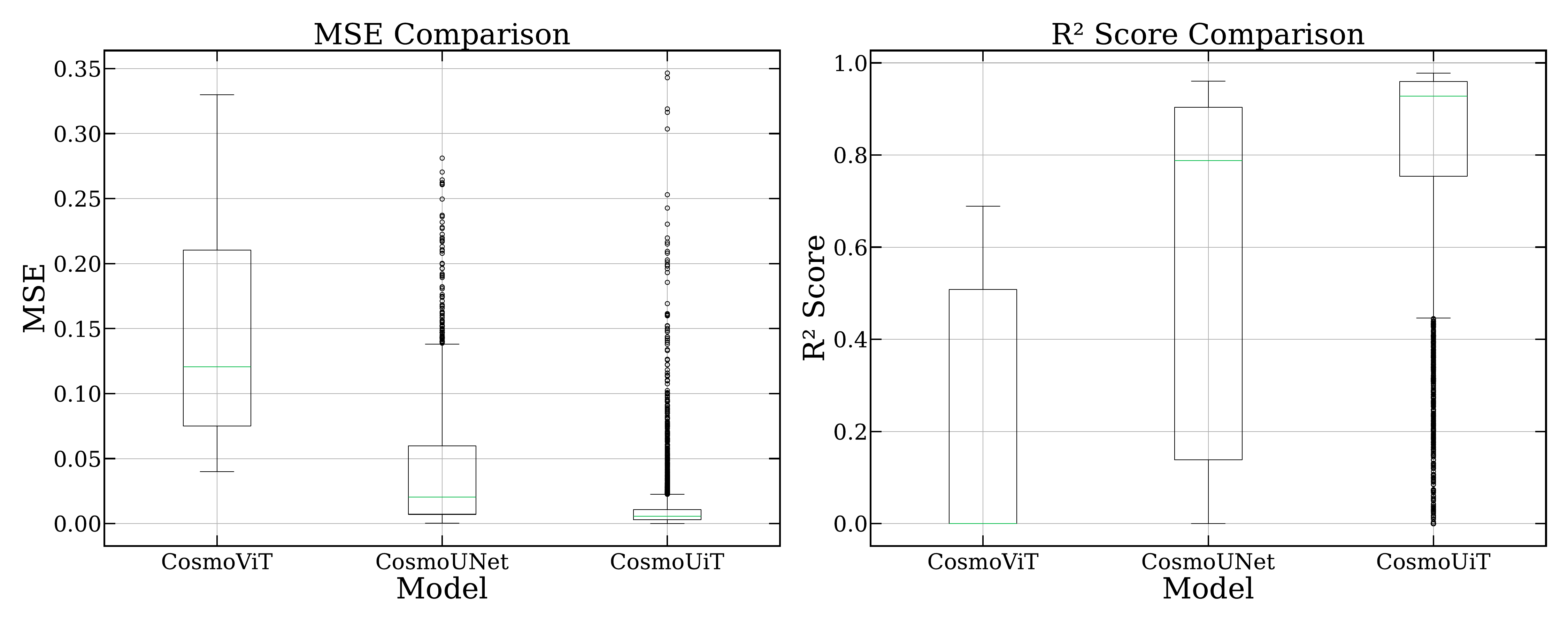}
    \caption{Comparison between $\text{R}^2$ scores of different model architectures.}
    \label{fig:metric_comparison}
\end{figure}

Figure \ref{fig:metric_comparison} presents boxplots comparing the distribution of MSE and R$^2$ scores for the three model architectures discussed above. These metrics are evaluated across multiple combinations of reionization parameters. In each boxplot, the blue horizontal line indicates the median of the distribution, the bottom and top edges of the box represent the first (Q1) and third (Q3) quartiles, respectively, and the whiskers extend to 1.5 times the interquartile range (IQR) from the quartiles. Outliers beyond this range are shown as individual points.

The MSE comparison on the left shows that \texttt{CosmoUiT48} yields the lowest median error and the narrowest interquartile range, indicating both high accuracy and low variability in its predictions. In contrast, \texttt{CosmoViT} exhibits the highest median MSE and a broad spread, reflecting poor and inconsistent performance. \texttt{CosmoUNet} performs better than \texttt{CosmoViT} but shows significant variability and a long tail of high-error outliers.

In cases where the neutral fraction is extremely low, the  $\text{R}^2$ score becomes highly negative due to the low variance in the true field, making it difficult to compare model performances. For clearer visual interpretation, negative R$^2$ values have been clipped to zero in Figure \ref{fig:metric_comparison}. The R$^2$ score comparison on the right panel of the Figure demonstrates the superior performance of \texttt{CosmoUiT48}, with a median close to $1$ and minimal dispersion, indicating consistent and accurate predictions across parameter combinations. \texttt{CosmoUNet} achieves moderately high median scores but exhibits greater variability, suggesting less stable generalization. \texttt{CosmoViT}, on the other hand, shows the weakest performance, with both lower median scores and a broader distribution.

These comparisons demonstrate that \texttt{CosmoUiT48} achieves the highest accuracy among the three models and generalizes more consistently across a broad range of reionization parameter combinations.

\section{Uncertainty Estimation} \label{sec:uncertainty_qantification}
Deep learning-based emulators are statistical approximations; therefore, their predictions are subject to error. If these errors are not properly accounted for in Bayesian inference pipelines, where the emulator is used as a model in the likelihood estimation, they may lead to biased estimates of reionization parameters. To address this issue, we explored multiple approaches for uncertainty estimation and evaluated them by measuring the correlation between the predicted uncertainty and the root mean squared error (RMSE). We then selected the method that produced the highest correlation as our final choice. Our strategy is to generate an ensemble of slightly different predictions for a fixed set of reionization parameters. From this ensemble, the pixel-wise mean is taken as the final prediction, while the pixel-wise standard deviation provides an estimate of the prediction uncertainty. The RMSE is then calculated by comparing the pixel-wise prediction with the ground truth obtained from the reference simulation. To produce the ensembles, we follow the data augmentation technique used in SegU-Net \citep{Bianco_2021}. Specifically, we apply 48 possible cube orientations (rotations and flips) and generate predictions for each. The predictions are then transformed back to the original orientation, allowing us to compute the mean field, uncertainty map, and pixel-wise RMSE as described above.  

\begin{figure}[htbp]
    \centering
    \includegraphics[width=0.98\linewidth]{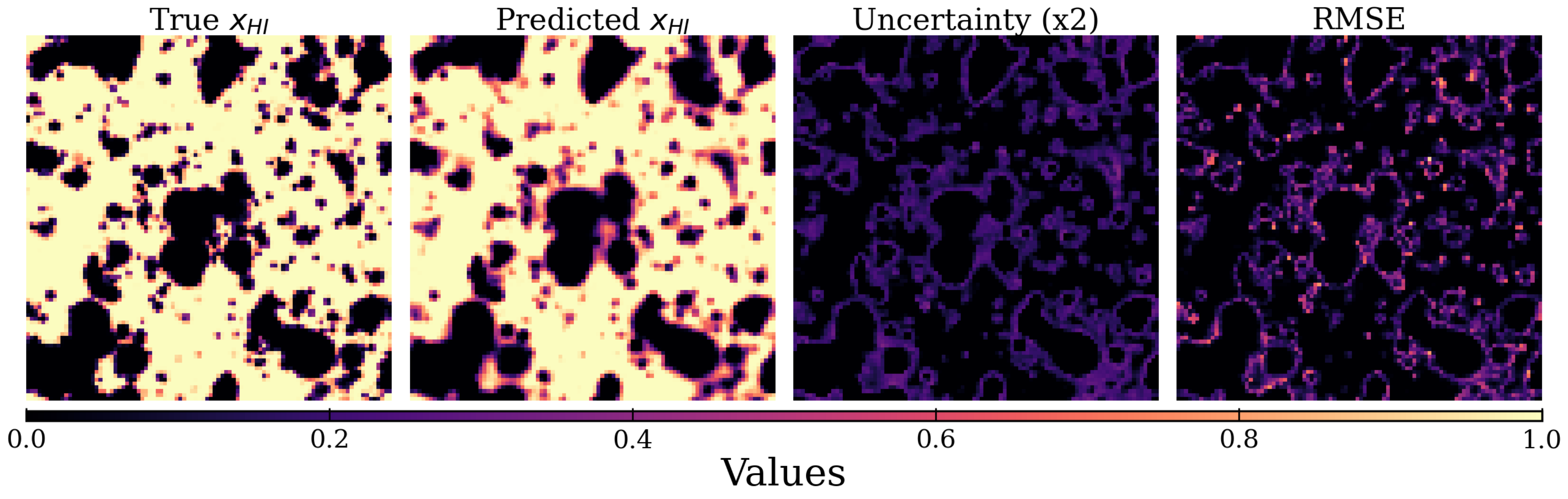}
    \caption{Uncertainty estimation via data augmentation: true and predicted neutral fraction field, associated uncertainty, and RMSE.}
\label{fig:uncertainty_estimation_using_data_augmentation}
\end{figure}

\begin{figure}[htbp]
    \centering
    \includegraphics[width=0.49\linewidth]{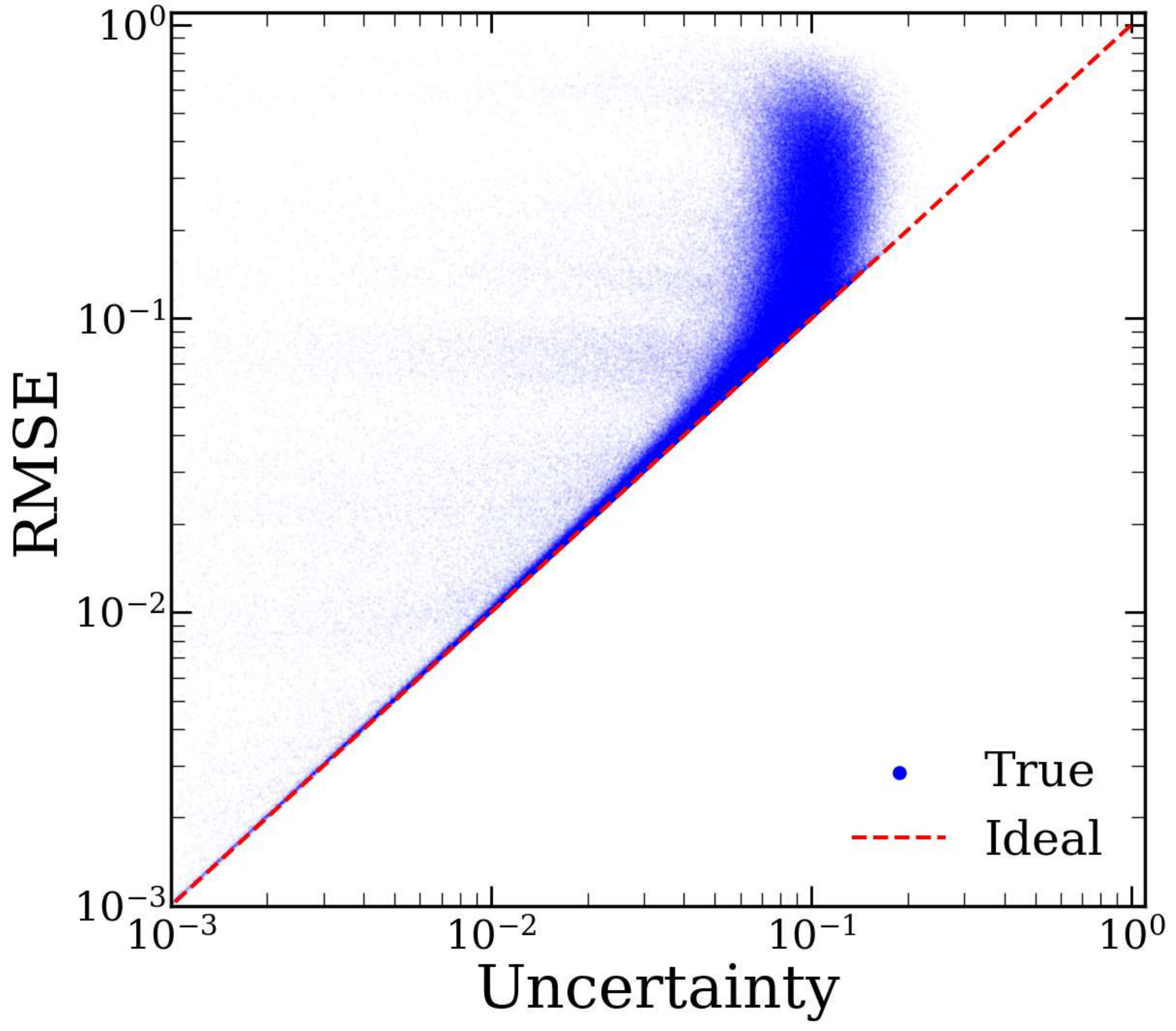}
    \caption{Pixel-wise correlation between predicted uncertainty and RMSE, obtained using data augmentation.}
  \label{fig:scatterplot_of_uncertainty}
    \end{figure}

Figure \ref{fig:uncertainty_estimation_using_data_augmentation} shows the comparison between the ground truth obtained via simulation, the mean prediction obtained from the prediction ensemble, pixel-wise uncertainty, and RMSE.
Figure~\ref{fig:scatterplot_of_uncertainty} presents the scatter plot of predicted uncertainty versus RMSE. A strong correlation is observed, with a coefficient of 0.83, which is significantly higher than that obtained with the other methods we tested. Although this represents a substantial improvement, further refinement may be achieved by incorporating Bayesian layers into the network, which we leave for future work. Once quantified, these uncertainties can be propagated through the inference pipeline to get more robust estimates of reionization parameters.



\bibliographystyle{JHEP}
\bibliography{biblio.bib}

\end{document}